\documentclass{acmtrans2m_cgg}
\acmVolume{V}
\acmNumber{N}
\acmYear{YY}
\acmMonth{M}

\newdef{definition}[theorem]{Definition}
\newdef{remark}[theorem]{Remark}
\newdef{notion}[theorem]{Notion}

\usepackage{amsfonts}
\usepackage{amsmath}
\usepackage{graphicx}

\title{A General Methodology for Designing Self-Organizing Systems}
\author{CARLOS GERSHENSON \\
 Vrije Universiteit Brussel}

\begin{abstract}
Our technologies complexify our environments. Thus, new technologies need to
deal with more and more complexity. Several efforts have been made to deal
with this complexity using the concept of self-organization. However, in order to promote its use and understanding, we
must first have a pragmatic understanding of complexity and
self-organization. This paper presents a conceptual framework for speaking
about self-organizing systems. The aim is to provide a methodology useful
for designing and controlling systems developed to solve complex problems.
First, practical notions of complexity and self-organization are given.
Then, starting from the agent metaphor, a conceptual framework is
presented. This provides formal ways of speaking about \textquotedblleft
satisfaction\textquotedblright\ of elements and systems. The main premise of
the methodology claims that reducing the \textquotedblleft
friction\textquotedblright\ or \textquotedblleft
interference\textquotedblright\ of \emph{interactions} between elements of a
system will result in a higher \textquotedblleft
satisfaction\textquotedblright\ of the system, i.e. better performance. The
methodology discusses different ways in which this can be achieved. A case
study on self-organizing traffic lights illustrates the ideas presented in
the paper.
\end{abstract}


            
            
\begin{document}
            
\begin{bottomstuff} 

Author's address: Krijgskundestraat 33 B-1160 Brussel, Belgium \newline
cgershen@vub.ac.be http://homepages.vub.ac.be/\symbol{126}cgershen
\end{bottomstuff}
            
\maketitle

\section{Introduction}

Over the last half a century, much research in different areas has employed
self-organizing systems to solve complex problems, e.g. \cite%
{Ashby1956,Beer1966,BonabeauEtAl1999,EngineeringSOS2004,ZambonelliRana2005}.
Recently, particular methodologies using the concepts of self-organization
have been proposed in different areas, such as software engineering \cite%
{WooldridgeEtAl2000,ZambonelliEtAl2003}, electrical engineering \cite%
{RamamoorthyEtAl1993}, and collaborative support \cite{JonesEtAl1994}.
However, there is as yet no general framework for constructing
self-organizing systems. Different vocabularies are used in different areas,
and with different goals. In this paper, I\ present an attempt to develop a
general methodology that will be useful for designing and controlling \emph{%
complex} systems \cite{Bar-Yam1997}. The proposed methodology, as with any
methodology, does not provide ready-made solutions to problems. Rather, it
provides a \emph{conceptual framework}, a \emph{language,} to assist the
solution of problems. Also, many current problem solutions can be \emph{%
described} as proposed. I am not suggesting new solutions, but an
alternative way of thinking about them.

As an example, many standardization efforts have been advanced in recent years, such as ontologies required for the Semantic Web \cite{Berners-LeeEtAl2001}, or FIPA standards. I am not insinuating that standards are not necessary. Without them engineering would be chaos. But as they are now, they cannot predict future requirements. They are developed with a static frame of mind. They are not adaptive. What this work suggests is a way of introducing the expectation of change into the development process to be able to cope with the unexpected beforehand, in problem domains where this is desired.

The paper is organized as follows: in the next section, notions of
complexity and self-organization are discussed. In Section \ref%
{secConceptualFw}, original concepts are presented. These will be used in
the Methodology, exposed in Section \ref{secMethodology}. In Section \ref%
{secSOTL}, a case study concerning self-organizing traffic lights is used to
illustrate the steps of the Methodology. Discussion and conclusions follow
in Sections \ref{secDiscussion} and \ref{secConclusions}.

\section{Complexity and Self-organization}

There is no general definition of \emph{complexity}, since the concept
achieves different meanings in different contexts \cite{Edmonds1999}. Still, we can say that a
system is complex if it consists of several \emph{interacting} elements \cite{Simon1996}, so
that the behavior of the system will be difficult to deduce from the
behavior of the parts. This occurs when there are many parts, and/or when
there are many interactions between the parts. Typical examples of complex
systems are a living cell, a society, an economy, an ecosystem, the
Internet, the weather, a brain, and a city. These all consist of numerous
elements whose interactions produce a global behavior that cannot be reduced
to the behavior of their separate components \cite{GershensonHeylighen2005}. For example, a cell is considered a living system, but the elements that conform it are not alive. The properties of life arise from the complex dynamical \emph{interactions} of the components. The properties of a system that are not present at the lower level (such as life), but are a product of the interactions of elements, are sometimes called \emph{emergent} \cite{Anderson1972}. Another example can be seen with gold: it has properties, such as temperature, malleability, conductivity, and color, that emerge from the interactions of the gold atoms, since atoms do not have these properties.

Even when there is no general definition or measure of complexity, a
relative \emph{notion} of complexity can be useful:

\newdef{notionComplexity}[theorem]{Notion}

\begin{notionComplexity}
The complexity of a
system scales with the number of its elements, the number of interactions
between them, the complexities of the elements, and the complexities of the
interactions \cite{Gershenson2002a}:\footnote{%
This can be confirmed mathematically in certain systems. As a general
example, random Boolean networks \cite%
{Kauffman1969,Kauffman1993,Gershenson2004c} show clearly that the complexity
of the network increases with the number of elements and the number of
interactions.} 
\end{notionComplexity}
\begin{equation}
C_{sys} \sim \#\overline{E} \#\overline{I} 
\sum_{j=0}^{\#\overline{E}}C_{e_{j}}
\sum_{k=0}^{\#\overline{I}}C_{i_{k}}
\label{eqComplexity}
\end{equation}

The complexity on an interaction $C_{i}$ can be measured as the number of different possible interactions two elements can have.\footnote{Certainly, the number of possible interactions for certain elements is impossible to enumerate or measure.}

The problem of a strict definition of complexity lies in the fact that there is no way of drawing a line between simple and complex systems independently of a context. For example, the \emph{dynamics} of a system can be simple (ordered), complex, or chaotic, having a complex structure. Cellular automata and random Boolean networks are a clear example of this, where moreover, the interactions of their components are quite simple. On the other hand, a \emph{structurally} simple system can have complex and chaotic dynamics. For this case, the damped pendulum is a common example.
Nevertheless, for practical purposes, the above notion will suffice, since it allows the comparison of the complexity of one system with another under a common frame of reference. Notice that the notion is recursive, so a basic level needs to be set contextually for comparing two systems.

The term \emph{self-organization} has been used in different areas with
different meanings, as is cybernetics \cite{vonFoerster1960,Ashby1962},
thermodynamics \cite{NicolisPrigogine1977}, biology \cite{CamazineEtAl2003},
mathematics \cite{Lendaris1964}, computing \cite{HeylighenGershenson2003},
information theory \cite{Shalizi2001}, synergetics \cite{Haken1981}, and
others \cite{SkarCoveney2003} (for a general overview, see \cite%
{Heylighen2003sos}). However, the use of the term is subtle, since any
dynamical system can be said to be self-organizing or not, depending partly
on the observer \cite{GershensonHeylighen2003a,Ashby1962}: If we decide to call a ``preferred" state or set of states (i.e. attractor) of a system ``organized", then the dynamics will lead to a self-organization of the system.

It is not necessary to enter into a philosophical debate on the theoretical
aspects of self-organization to work with it, so a practical notion will
suffice: 

\newdef{notionSelfOrg}[theorem]{Notion}

\begin{notionSelfOrg}

A system \emph{described} as self-organizing is one in which
elements \emph{interact} in order to achieve \emph{dynamically} a global function or behavior.

\end{notionSelfOrg}

This function or behavior is not imposed by one single or a few elements,
nor determined hierarchically. It is achieved \emph{autonomously} as the elements
interact with one another. These interactions produce feedbacks that
regulate the system. All the previously mentioned examples of complex
systems fulfill the definition of self-organization. More precisely, the
question can be formulated as follows: \emph{when is it useful to describe a
system as self-organizing?} This will be when the system or environment is
very dynamic and/or unpredictable. If we want the system to solve a problem,
it is useful to describe a complex system as self-organizing when the
``solution" is not known beforehand and/or is changing constantly. Then, the
solution is dynamically strived for by the elements of the system. In this
way, systems can adapt quickly to unforeseen changes as elements interact
locally. In theory, a centralized approach could also solve the problem, but
in practice such an approach would require too much time to compute the
solution and would not be able to keep the pace with the changes in the
system and its environment.

In engineering, a self-organizing system would be one in which elements are
designed in order to solve \emph{dynamically} a problem or perform a function at the system level. Thus, the elements need to divide, but also integrate, the problem. For example, the parts of a car are designed to perform a function at the system level: to drive. However, the parts of a (normal) car do not change their behavior in time, so it might be redundant to call a car self-organizing. On the other hand, a swarm of robots \cite{DorigoEtAl2004} will be conveniently described as self-organizing, since each element of the swarm can change its behavior depending on the current situation.
It should be noted that all engineered self-organizing systems are to a certain degree \emph{autonomous}, since part of their actual behavior will not be determined by a designer.

In order to understand self-organizing systems, two or more \emph{levels of
abstraction} \cite{Gershenson2002a} should be considered: elements (lower
level) organize in a system (higher level), which can in turn organize with
other systems to form a larger system (even higher level). The understanding
of the system's behavior will come from the relations observed between the
descriptions at different levels. Note that the levels, and therefore also
the terminology, can change according to the interests of the observer. For
example, in some circumstances, it might be useful to refer to cells as
elements (e.g. bacterial colonies); in others, as systems (e.g. genetic
regulation); and in others still, as systems coordinating with other systems
(e.g. morphogenesis).

A system can cope with an unpredictable environment \emph{autonomously} using different but closely related approaches:
\begin{itemize}
    \item \textbf{Adaptation} (learning, evolution) \cite{Holland1995}. The system changes its behavior to cope with the change.
	\item \textbf{Anticipation} (cognition) \cite{Rosen1985}. The system predicts a change to cope with, and adjusts its behavior accordingly. This is a special case of adaptation, where the system does not require to experience a situation before responding to it.
    \item \textbf{Robustness} \cite{vonNeumann1956,Jen2005}. A system is robust if it continues to function in the face of perturbations 
\cite{Wagner2005}. This can be achieved with modularity
\cite{Simon1996,Watson2002}, degeneracy
\cite{FernandezSole2003}, distributed robustness
\cite{Wagner2004}, or redundancy \cite{GershensonEtAl2006}.
\end{itemize} 
Successful self-organizing systems will use combinations of the these approaches to maintain their integrity in a changing and unexpected environment. Adaptation will enable the system to modify itself to ``fit" better within the environment. Robustness will allow the system to withstand changes without losing its function or purpose, and thus allowing it to adapt. Anticipation will prepare the system for changes before these occur, adapting the system without it being perturbed. We can see that all of them should be taken into account while engineering self-organizing systems.

In the following section, further concepts will be introduced that will be
necessary to apply the methodology .

\section{The Conceptual Framework}\label{secConceptualFw}

Elements of a complex system interact with each other. The actions of one
element therefore affect other elements, directly or indirectly. For
example, an animal can kill another animal directly, or indirectly cause its
starvation by consuming its resources. These interactions can have negative,
neutral, or positive effects on the system \cite{HeylighenCampbell1995}.

Now, intuitively thinking, it may be that the ``smoothening" of local
interactions, i.e. the minimization of ``interferences" or ``friction" will
lead to global improvement. But is this always the case? To answer this
question, the terminology of multi-agent systems \cite%
{Maes1994,WooldridgeJennings1995,Wooldridge2002,Schweitzer2003}\ can be
used. We can say that: 

\newdef{notionAgent}[theorem]{Notion}

\begin{notionAgent}

An agent is a description of an entity that \emph{acts} on its environment.
\end{notionAgent}

Examples of this can be a trader acting on a market, a school of fish acting on a coral reef, or a computer acting on a network.
Thus, every element, and every system, can be seen as agents with \emph{goals}
and behaviors thriving to reach those goals. The behavior of agents can
affect (positively, negatively, or neutrally) the fulfillment of the goals of
other agents, thereby establishing a relation. The \emph{satisfaction} or
fulfillment of the goals of an agent can be represented using a variable $%
\sigma \in \lbrack 0,1]$.\footnote{%
In some cases, $\sigma $ could be seen as a ``fitness" \cite%
{HeylighenCampbell1995}. However, in most genetic algorithms \cite%
{Mitchell1996} a fitness function is imposed from the outside, whereas $%
\sigma $ is a property of the agents, that can change with time.} Relating
this to the higher level, the satisfaction of a system $\sigma _{sys}$ can
be recursively represented as a function $f:
\mathbb{R}
\rightarrow \lbrack 0..1]$ of the satisfaction of the\ $n$ elements
conforming it:

\begin{equation}
\sigma _{sys}=f\left( \sigma _{1},\sigma _{2},...,\sigma
_{n},w_{0},w_{1},w_{2},...,w_{n}\right)  \label{eqSigmaSys}
\end{equation}

where $w_{0}$ is a bias and the other weights determine the importance given
to each $\sigma _{i}$. If the system is homogeneous, then $f$ will be the
weighted sum of $\sigma _{i}$, $w_{i}=\frac{1}{n}\forall i\neq 0$, $w_{0}=0$%
. Note that this would be very similar to the activation function used in many artificial neural networks \cite{Rojas1996}.
 For heterogenous systems, $f$ may be a nonlinear function. Nevertheless, the weights $w_{i}$'s are determined \emph{%
tautologically} by the importance of the $\sigma $ of each element to the
satisfaction of the system. Thus, it is a useful tautology to say that
maximizing individual $\sigma $'s, adjusting individual behaviors (and thus
relations), will maximize $\sigma _{sys}$. If several elements decrease $%
\sigma _{sys}$ as they increase their $\sigma $, we would not consider them
as part of the system. It is important to note that this is independent of the potential nonlinearity of $f$. An example can be seen with the immune system. It categorizes molecules and micro-organisms as akin or alien \cite%
{VazVarela1978}. If they are considered as alien, they are attacked.
Auto-immune diseases arise when this categorization is erroneous, and the
immune system attacks vital elements of the organism. On the other hand, if
pathogens are considered as part of the body, they are not attacked. Another
example is provided by cancer. Carcinogenic cells can be seen as ``rebel",
and no longer part of the body, since their goals differ from the goal of
the organism. Healthy cells are described easily as part of an organism. But
when they turn carcinogenic, they can better be described as parasitic. The
tautology is also useful because it gives a general mathematical
representation for system satisfaction, which is independent of a particular
system.

A reductionist approach would assume that maximizing the satisfaction of the
elements of a system would also maximize the satisfaction of the system.
However, this is not always the case, since some elements can ``take
advantage" of other elements. Thus, we need to concentrate \emph{also} on
the interactions of the elements.

If the model of a system considers more than two levels, then the $\sigma $
of higher levels will be recursively determined by the $\sigma $'s of lower
levels. However, the $f$'s most probably will be very different on each
level.

Certainly, an important question remains: how do we determine the function $%
f $ and the weights $w_{i}$'s? To this question there is no complete answer.
One option would be to approximate $f$ numerically \cite{DeWolfEtAl2005}. An explicit $f$ may be difficult to find, but an approximation can be very useful.
Another method consists of \emph{lesioning} the system\footnote{%
This method has been used widely to detect functions in complex systems such
as genetic regulatory networks and nervous systems.}: removing or altering
elements of the system, and observing the effect on $\sigma _{sys}$. Through
analyzing the effects of different lesions, the function $f$ can be
reconstructed and the weights $w_{i}$'s obtained. If a small change $\Delta
\sigma _{i}$ in any $\sigma _{i}$ produces a change $\Delta \sigma
_{sys}\geq \Delta \sigma _{i}$, the system can be said to be \emph{fragile}.

What could then be done to maximize $\sigma _{sys}$? How can we relate the $%
\sigma _{i}$'s and avoid conflicts between elements? This is not an obvious
task, for it implies bounding the agents' behaviors that reduce other $%
\sigma _{i}$'s, while preserving their functionality. Not only should the
interference or friction between elements be minimized, but the synergy \cite%
{Haken1981}\ or ``positive interference" should also be promoted. Dealing
with complex systems, it is not feasible to tell each element what to do or
how to do it, but their behaviors need to be constrained or modified so
that their goals\ will be reached, blocking the goals of other elements as
little as possible. These constraints can be called \emph{mediators} \cite%
{Michod2003,Heylighen2003}. They can be imposed from the top down, developed from the
bottom up, be part of the environment, or be embedded as an \emph{aspect} 
\cite[Ch. 3]{tenHaafEtAl2002} of the system. An example can be found in city
traffic: traffic lights, signals and rules mediate among drivers, trying to
minimize their conflicts, which result from the competition for limited
resources, i.e. space to drive through. The role of a mediator is to
arbitrate among the elements of a system, to minimize interferences and
frictions and maximize synergy. Therefore, the efficiency of the mediator
can be measured directly using $\sigma _{sys}$. Individually, we can measure
the ``friction" $\phi _{i}\in \lbrack -1,1]$ that agent $i$ causes in the
rest of the system, relating the change in satisfaction $\Delta \sigma _{i}$
of element $i$\ and the change in satisfaction of the system $\Delta \sigma
_{sys}$:

\begin{equation}
\phi _{i}=\frac{-\Delta \sigma _{i}-\Delta \sigma _{sys}\left( n-1\right) }{n%
}.  \label{eqFriction}
\end{equation}

Friction occurs when the increase of satisfaction of one element causes a
decrease in the satisfaction of some other elements that is greater than the
increase. Note that $\phi _{i}=0$ does imply that there is no conflict,
since one agent can ``get" the satisfaction proportionally to the ``loss" of
satisfaction of (an)other agent(s). Negative friction would imply synergy,
e.g. when $\Delta \sigma _{i}\geq 0$ while other elements also increase
their $\sigma $. The role of a mediator would be to maximize $\sigma _{sys}$
by minimizing $\phi _{i}$'s. With this approach, friction can be seen as a
type of \emph{interaction} between elements.

Thus, the problem can be put in a different way: how can we
find/develop/evolve efficient mediators for a given system? One answer to
this question is the methodology proposed in this paper. The answer will not
be complete, since we cannot have precise knowledge of $f$ for large
evolving complex systems. This is because the evolution of the system will
change its own $f$ \cite{Kauffman2000}, and the relationships among
different $\sigma _{i}$'s. Therefore, predictions cannot be complete.
However, the methodology proposes to follow steps to increase the
understanding (and consequently the control) of the system and the relations
between its elements. The goal is to identify conflicts and diminish them
without creating new ones. This will increase the $\sigma _{i}$'s and thus $%
\sigma _{sys}$. The precision of $f$ is not so relevant if this is achieved.

It should be noted that the timescale chosen for measuring $\Delta \sigma
_{i}$ is very important, since at short timescales the satisfaction can
decrease, while on the long run it will increase. In other words, there can
be a short term ``sacrifice" to harvest a long term ``reward". If the
timescale is too small, a system might get stuck in a ``local optimum", since
all possible actions would decrease its satisfaction on the short term. But
in some cases the long term benefit should be considered for maximization. A
way of measuring the slow change of $\sigma _{i}$ would be with its integral
over time for a certain interval $\Delta t$:

\begin{equation}
\int_{t}^{t+\Delta t}\sigma _{i}dt.  \label{eqIntegralSigma}
\end{equation}

Another way of dealing with the local optima is to use neutral changes to
explore alternative solutions \cite{Kimura1983}.

Before going into further detail, it is worth noting that this is not a
reductionist approach. Smoothing out local interactions will not provide
straightforward clues as to what will occur at the higher level. Therefore,
the system should be observed at both levels: making local and global
changes, observing local and global behaviors, and analyzing how one
affects the other.

Concurrently, the \textit{dependence} $\epsilon $ $\in \lbrack -1,1]$ of an
element to the system can be measured by calculating the difference of the
satisfaction $\sigma _{i}$ when the element interacts within the system and
its satisfaction $\widetilde{\sigma _{i}}$ when the element is isolated.

\begin{equation}
\epsilon =\sigma _{i}-\widetilde{\sigma _{i}}.  \label{eqDependence}
\end{equation}

In this way, full dependence is given when the satisfaction of the element
within the system $\sigma _{i}$ is maximal and its satisfaction $\widetilde{\sigma _{i}}$ is minimal when the element is isolated. A negative $\epsilon 
$ would imply that the element would be more satisfied on its own and is
actually ``enslaved" by the system. Now we can use the dependences of the
elements to a system to measure the \textit{integration} $\tau $ $\in
\lbrack -1,1]$ of a system, which can be seen also as a gradual measure of a
meta-system transition (MST) \cite{Turchin1977}.

\begin{equation}
\tau =\frac{1}{n}\sum\limits_{i=1}^{n}\epsilon _{i}.
\label{eqIntegration-MST}
\end{equation}

A MST is a gradual process, but it will be complete when elements are not
able to reach their goals on their own, i.e. $\overline{\sigma _{i}}%
\rightarrow 0$. Examples include cells in multi-cellular organisms and
mitochondria in eukaryotes.

In an evolutionary process, natural (multilevel \cite{Michod1997,Lenaerts2003}) selection will tend to increase $\tau $
because this implies higher satisfaction both for the system and its
elements (systems with a negative $\tau $ are not viable). Relations and
mediators that contribute to this process will be selected, since higher $%
\sigma $'s imply more chances of survival and reproduction. Human designers
and engineers also select relations and mediators that increase the $\sigma $%
's of elements and systems. Therefore, we can see that evolution will tend,
in the long run, towards synergetic relationships \cite{Corning2003}, even
if resources are scarce.

In the next section, the steps suggested for developing a self-organizing
system are presented, using the concepts described in this section.

\section{The Methodology}\label{secMethodology}

\begin{figure*}[tp]
\begin{center}
\includegraphics[
width= 12.1144cm, height=6.3395cm]
{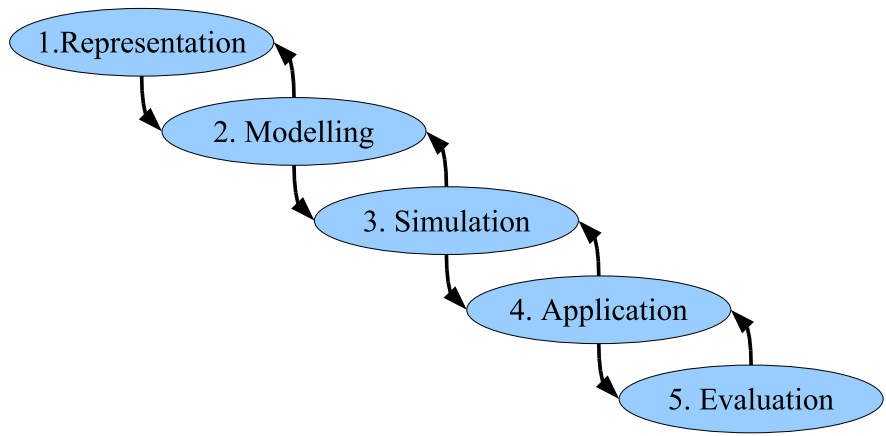}
\end{center}
\caption{Diagram relating different stages of methodology}
\label{diagram}
\end{figure*}

The proposed methodology meets the requirements of a system, i.e. what the
system should do, and enables the designer to produce a system that fulfills
the requirements. The methodology includes the following steps:
Representation, Modeling, Simulation, Application, and Evaluation, which
will be exposed in the following subsections. Figure \ref{diagram} presents
these steps. These steps should not necessarily be followed one by one,
since the stages merge with each other. There is also backtracking, when the
designer needs to return to an earlier stage for reconsideration before finishing a cycle.

This methodology should not be seen as a recipe that provides ready-made
solutions, but rather as a guideline to direct the search for them. The stages proposed are not new, and similar to those proposed by iterative and incremental development methodologies. Still, it should be noted that the active feedback between stages within each iteration can help in the design of systems ready to face uncertainties in complex problem domains. The novelty of the methodology lies in the \emph{vocabulary} used to describe self-organizing systems.

\subsection{Representation}

The goal of this step is to develop a \emph{specification} (which might be
tentative) of the components of the system.

The designer should always remember the distinction between model and
modeled. A model is an abstraction/description of a ``real" system. Still,
there can be several descriptions of the same system \cite%
{Gershenson2002a,GershensonHeylighen2005}, and we cannot say that one is
better than another independently of a context.

There are many possible representations of a system. According to the \emph{constraints} and \emph{requirements}, which may be incomplete, the designer
should choose an appropriate vocabulary (metaphors to speak about the
system), abstraction levels, granularity, variables, and interactions that
need to be taken into account. Certainly, these will also depend on the
experience of the designer. The choice between different approaches can
depend more on the expertise of the designer than on the benefits of the
approaches.

Even when there is a wide diversity of possible systems, a general approach
for developing a Representation can be abstracted. The designer should try
to divide a system into elements by identifying semi-independent modules,
with internal goals and dynamics, and with few interactions with their
environment. Since interactions in a model will increase the complexity of
the model, we should group ``clusters" of interacting variables into
elements, and then study a minimal number of interactions between elements.
The first constraints that help us are space and time. It is useful to group
variables that are close to each other (i.e. interacting constantly) and
consider them as elements that relate to other elements in occasional
interactions. Multiscale analysis \cite{Bar-Yam2005} is a promising method for identifying levels and variables useful in a Representation.
Since the proposed methodology considers elements as agents,
another useful criterion for delimiting them is the identification of goals.
These will be useful in the Modeling to measure the satisfaction $\sigma $
of the elements. We can look at genes as an example: groups of nucleotides
co-occur and interact with other groups and with proteins. Genes are
identified by observing nucleotides that keep close together and act
together to perform a function. The fulfillment of this function can be seen
as a goal of the gene. Dividing the system into modules also divides the
problem it needs to solve, so a complex task will be able to be processed in
parallel by different modules. Certainly, the integration of the ``solutions"
given by each module arises as a new problem. Nevertheless, modularity in a
system also increases its robustness and adaptability \cite{Simon1996,Watson2002,FernandezSole2003}.

The representation should consider at least two levels of abstraction, but
if there are many variables and interactions in the system, more levels can
be contemplated. Since elements and systems can be seen as agents, we can
refer to all of them as $x$-agents, where $x$ denotes the level of
abstraction relative to the simplest elements. For example, a three-layered
abstraction would contemplate elements (0-agents) forming systems that are
elements (subsystems, 1-agents) of a greater system (meta-system, 2-agents).
If we are interested in modeling a research institute, 0-agents would be
researchers, 1-agents would be research groups, and the research institute
would be a 2-agent. Each of these have goals and satisfactions ($\sigma ^{x}$) that can be described and interrelated. For engineering purposes, the
satisfaction of the highest level, i.e. the satisfaction of the system that
is being designed, will be determined by the tasks expected from it. If
these are fulfilled, then it can be said that the system is ``satisfied".
Thus, the designer should concentrate on engineering elements that will
strive to reach this satisfaction.

If there are few elements or interactions in the Representation, there will
be low complexity, and therefore stable dynamics. The system might be better
described using traditional approaches, since the current approach might
prove redundant. A large variety of elements and/or interactions might imply a high complexity. Then, the Representation should be revised before entering the Modeling stage.

\subsection{Modeling}

In science, models should ideally be as simple as possible, and predict as
much as possible \cite{Shalizi2001}. These models will provide a better
understanding of a phenomenon than complicated models. Therefore, a good
model requires a good Representation. The ``elegance" of the model will
depend very much on the metaphors we use to speak about the system. If the
model turns out to be cumbersome, the Representation should be revised.

The Modeling should specify a Control\emph{\ }mechanism that will ensure
that the system does what it is required to do. Since we are interested in
self-organizing systems, the Control will be \emph{internal} and \emph{%
distributed}. If the problem is too complex, it can be divided into
different subproblems. The Modeling should also consider different
trade-offs for the system.

\subsubsection{Control mechanism}

The Control mechanism can be seen as a \emph{mediator} \cite{Heylighen2003}
ensuring the proper interaction of the elements of the system, and one that
should produce the desired performance. However, one cannot have a strict
control over a self-organizing system. Rather, the system should be \emph{%
steered} \cite{Wiener1948}. In a sense, self-organizing systems are like
teenagers: they cannot be tightly controlled since they have their own
goals. We can only attempt to steer their actions, trying to keep their
internal variables under certain boundaries, so that the systems/teenagers
do not ``break" (in Ashby's sense \cite{Ashby1947}).

To develop a Control, the designer should find aspect systems, subsystems,
or constraints that will prevent the negative interferences between elements
(friction) and promote positive interferences (synergy). In other words, the
designer should search for ways of minimizing frictions $\phi _{i}$'s that
will result in maximization of the global satisfaction $\sigma _{sys}$. The
performance of different mediators can be measured using equation (\ref%
{eqSigmaSys}).

The Control mechanism should be \emph{adaptive}. Since the system is dynamic
and there are several interactions within the system and with its
environment, the Control mechanism should be able to cope with the changes
within and outside the system, in other words, \emph{robust}. An adaptive Control will be efficient in more
contexts than a static one. In other words, the Control should be \emph{%
active} in the search of solutions. A static Control will not be able to
cope with the complexity of the system. There are several methods for
developing an adaptive Control, e.g. \cite{SastryBodson1994}. But these
should be applied in a distributed way, in an attempt to reduce friction and
promote synergy.

Different methods for reducing friction in a system can be identified. In
the following cases, an agent A negatively affected by the behavior of an
agent B will be considered\footnote{%
Even when equation \ref{eqFriction} relates the satisfaction of an element
to the satisfaction of the system, this can also be used for the relation
between satisfactions of two elements, when $\Delta \sigma _{i}=0$ for all
other elements.}:

\begin{itemize}
\item \textbf{Tolerance}. This can be seen as the acceptance of others and
their goals. A can tolerate B by modifying itself to reduce the friction
caused by B, and therefore increase $\sigma _{A}$. This can be done by
moving to another location, finding more resources, or making internal
changes.

\item \textbf{Courtesy}. This would be the opposite case to Tolerance. B
should modify its behavior not to reduce $\sigma _{A}$.

\item \textbf{Compromise}. A combination of Courtesy and Tolerance: both
agents A and B should modify their behaviors to reduce the friction. This
is a good alternative when both elements cause friction to each other. This
will be common when A and B are similar, as in homogeneous systems.

\item \textbf{Imposition}. This could be seen as forced Courtesy. The
behavior of B could be changed by force. The Control could achieve this by
constraining B or imposing internal changes.

\item \textbf{Eradication}. As a special case of Imposition, B can be
eradicated. This certainly would decrease $\sigma _{B}$, but can be an
alternative when either $\sigma _{B}$ does not contribute much to $\sigma
_{sys}$, or when the friction caused by B in the rest of the system is very
high.

\item \textbf{Apoptosis}. B can ``commit suicide". This would be a special
case of Courtesy, where B would destroy itself for the sake of the system.
\end{itemize}

The difference between Compromise/Apoptosis and Imposition/Eradication is
that in the former cases, change is triggered by the agent itself, whereas
in the latter the change is imposed from the ``outside" by a mediator.
Tolerance and Compromise could be generated by an agent or by a mediator.

Different methods for reducing friction can be used for different problems.
A good Control will select those in which other $\sigma $'s are not reduced
more than the gain in $\sigma $'s. The choice of the method will also depend
on the importance of different elements for the system. Since more important
elements contribute more to $\sigma _{sys}$, these elements can be given
preference by the Control in some cases.

Different methods for increasing synergy can also be identified. These will
consist of taking measures to increase $\sigma _{sys}$, even if some $\sigma 
$'s are reduced:

\begin{itemize}
\item \textbf{Cooperation}. Two or more agents adapt their behavior for the
benefit of the whole. This might or might not reduce some $\sigma $'s.

\item \textbf{Individualism}. An agent can choose to increase its $\sigma $
if it increases $\sigma _{sys}$. A mediator should prevent increases in $%
\sigma $'s if these reduce $\sigma _{sys}$, i.e. friction.

\item \textbf{Altruism}. An agent can choose to sacrifice an increase of its 
$\sigma $ or to reduce its $\sigma $ in order to increase $\sigma _{sys}$.
This would make sense only if the \emph{relative} increase of $\sigma _{sys}$
is greater than the decrease of the $\sigma $ of the altruistic agent. A
mediator should prevent wasted altruism.

\item \textbf{Exploitation}. This would be forced Altruism: an agent is
driven to reduce its $\sigma $ to increase $\sigma _{sys}$.
\end{itemize}

A common way of reducing friction is to enable agents to learn via reinforcement   \cite{KaelblingEtAl1996}. With this method, agents tend to repeat actions that bring them satisfaction and avoid the ones that reduce it. Evolutionary approaches, such as genetic algorithms \cite{Mitchell1996}, can also reduce friction and promote synergy. However, these tend to be ``blind", in the sense that variations are made randomly, and only their effects are evaluated. With the criteria presented above, the search for solutions can be guided with a certain aim. However, if the relationship between the satisfaction of the elements and the satisfaction of the system is too obscure, ``blind" methods remain a good alternative.

In general, the Control should explore different alternatives, trying to
constantly increase $\sigma _{sys}$ by minimizing friction and maximizing
synergy. This is a constant process, since a self-organizing system is in a
dynamic environment, producing ``solutions" for the current situation. Note that a mediator will not necessarily maximize the satisfaction of the agents. However, it should try to do so for the system.

\subsubsection{Dividing the problem}

If the system is to deal with many parameters, then it can be seen as a 
\emph{cognitive} system \cite{Gershenson2004}. It must ``know" or ``anticipate" what to do
according to the current situation and previous history. Thus, the main
problem, i.e. what the elements should do, could be divided into the
problems of communication, cooperation, and coordination \cite%
{GershensonHeylighen2004}.

For a system to self-organize, its elements need to \emph{communicate}: they
need to ``understand" what other elements, or mediators, ``want" to tell them.
This is easy if the interactions are simple: sensors can give \emph{meaning}
to the behaviors of other elements. But as interactions turn more complex,
the \emph{cognition} \cite{Gershenson2004} required by the elements should
also be increased. New meanings can be learned \cite{Steels1998,DeJong2000}\
to adapt to the changing conditions. These can be represented as ``concepts" 
\cite{Gardenfors2000}, or encoded, e.g., in the weights of a learning neural
network \cite{Rojas1996}. The precise implementation and philosophical interpretations are
not relevant if the outcome is the desired one.

The problem of \emph{cooperation} has been widely studied using game theory 
\cite{Axelrod1984}. There are several ways of promoting cooperation,
especially if the system is designed. To mention mention only two of them:
the use of tags \cite{RioloEtAl2001,HalesEdmonds2003} and multiple levels of
selection \cite{Michod1997,Lenaerts2003} have proven to yield cooperative behavior.
This will certainly reduce friction and therefore increase $\sigma _{sys}$.

Elements of a system should \emph{coordinate} while reducing friction, not
to obstruct each other. An important aspect of coordination is the \emph{%
division of labour}. This can promote synergy, since different elements can
specialize in what they are good at and \emph{trust\footnote{%
Trust is also important for communication and cooperation.}} others to do
what they are good at \cite{Gaines1994,DiMarzoSerugendo2004}. This process
will yield a higher $\sigma _{sys}$ compared to the case when every element
is meant to perform all functions independently of how well each element
performs each function. A good Control will promote division of labour by
mediating a balance between \emph{specialization} and \emph{integration}:
elements should devote more time doing what they are best at, but should
still take into account the rest of the system. Another aspect of
coordination is the \emph{workflow}: if some tasks are prerequisites of
other tasks, a mediator should synchronize the agents to minimize waiting
times.

\subsubsection{Trade-offs}

A system needs to be able to cope with the complexity of its domain to
achieve its goals. There are several trade-offs that can be identified to
reach a balance and cope better with this complexity:

\begin{itemize}
\item \textbf{Complexity of Elements/Interactions}. The complexity of the
system required to cope with the complexity of its domain can be tackled at
one end of the spectrum by complex elements with few/simple interactions, and at
the other by simple elements with several/complex interactions.

\item \textbf{Quality/Quantity}. A system can consist at one extreme of a
few complex elements, and at the other of several simple elements.

\item \textbf{Economy/Redundancy}. Solving a problem with as few elements as
possible is economical. But a minimal system is very fragile. Redundancy is
one way of favoring the \emph{robustness} of the system \cite%
{vonNeumann1966,FernandezSole2003,Wagner2004,GershensonEtAl2006}. Still, too much redundancy
can also reduce the speed of adaptation and increase costs for maintaining
the system.

\item \textbf{Homogeneity/Heterogeneity}. A homogeneous system will be
easier to understand and control. A heterogenous system will be able to cope
with more complexity with less elements, and will be able to adapt more
quickly to sudden changes. If there is a system of ten agents each able to
solve ten tasks, a homogeneous system will be able to solve more than ten
tasks robustly. A fully heterogeneous system would be able to solve more
than a hundred tasks, but it would be fragile if one agent failed. Heterogeneity also brings diversity, that can accelerate the speed of exploration, adaptation, and evolution, since different solutions can be sought in parallel. The diversity is also related to the amount of variety of perturbations that the system can cope with \cite{Ashby1956}, i.e. robustness.

\item \textbf{System/Context}. The processing and storage of information can
be carried out internally by the system, or the system can exploit its
environment ``throwing" complexity into it, i.e. allow it to store or process
information \cite{GershensonEtAl2003a}.

\item \textbf{Ability/Clarity}. A powerful system will solve a number of
complex problems, but it will not be very useful if the functioning of the
system cannot be understood. Designers should be able to understand the
system in order to be able to control it \cite{Schweitzer2003}.

\item \textbf{Generality/Particularity}. An abstract Modeling will enable
the designer to apply the Modeling in different contexts. However,
particular details should be considered to make the implementation feasible.
\end{itemize}

There are only very relative ways of measuring the above mentioned
trade-offs. However, they should be kept in mind during the development of
the system.

In a particular system, the trade-offs will become clearer once the Simulation is
underway. They can then be reconsidered and the Modeling updated.

\subsection{Simulation}

The aim here is to build computer simulation(s) that implement the model(s)
developed in the Modeling stage, and test different scenarios and mediator
strategies.

The Simulation development should proceed in stages: from abstract to
particular. First, an abstract scenario should be used to test the main
concepts developed during the Modeling. Only when these are tested and
refined, should details be included in the Simulation. This is because
particular details take time to develop, and there is no sense in investing
before knowing wether the Modeling is on the right track. Details can also
influence the result of the Simulation, so they should be put off until a
time when the main mechanisms are understood.

The Simulation should compare the proposed solutions with traditional
approaches. This is to be sure that applying self-organization in the system
brings any benefit. Ideally, the designer should develop more than one
Control to test in the simulation. A rock-scissors-paper situation could
arise, where no Control is best in all situations (also considering classic
controls). The designer can then adjust or combine the Controls, and then
compare again in the Simulation.

Experiments conducted with the aid of the Simulation should go from simple
to extensive. Simple experiments will show proof of concepts, and their
results can be used to improve the Modeling. Once this is robust, extensive
studies should be made to be certain of the performance of the system under
different conditions.

Based on the Simulation results and insights, the Modeling and
Representation can be improved. A Simulation should be mature before taking
the implementation into the real world.

\subsection{Application}

The role of this stage is basically to use the developed and tested model(s)
in a real system. If this is a software system, the transition will not be
so difficult. On the other hand, the transition to a real system can expose
artifacts of a naive Simulation. A useful way to develop robust Simulations
consists in adding some noise into the system \cite{Jakobi1997}.

Good theoretical solutions can be very difficult/expensive/impossible to
implement (e.g. if they involve instantaneous access to information, mind
reading, teleportation, etc.). The feasibility of the Application should be
taken into account during the whole design process. In other words, the
designer should have an \emph{implementation bias} in all the Methodology
stages. If the proposed system turned out to be too expensive or
complicated, all the designer's efforts would be fruitless. If the system is
developed for a client, there should be feedback between developers and
clients during the whole process \cite{Cotton1996} to avoid client
dissatisfaction once the system is implemented. The \emph{legacy} of
previous systems should also be considered for the design \cite%
{ValckenaersEtAl2003}: if the current implementation is to be modified but
not completely replaced, the designer is limited by the capabilities of the
old system.

Constraints permitting, a pilot study should be made before engaging in a
full Application, to detect incongruences and unexpected issues between the
Simulation or Modeling stages and the Application. With the results of this
pilot study, the Simulation, Modeling, and Representation can be fine-tuned.

\subsection{Evaluation}

Once the Application is underway, the performance of new system should be
measured and compared with the performances of the previous system(s).

Constraints permitting, efforts should be continued to try to improve the
system, since the requirements it has to meet will certainly change with
time (e.g. changes of demand, capacity, etc.). The system will be more
adaptive if it does not consider its solution as the best once and for all,
and is able to change itself according to its performance and the changing
requirements.

\subsection{Notes on the methodology}

\begin{itemize}
\item All returning arrows in the Figure \ref{diagram} are given because it
is not possible to predict the outcome of strategies before they have been
tried out. All information and eventualities cannot be abstracted, nor
emergent properties predicted before they have been observed. Emergent
properties are \textit{a posteriori}.

\item The proposed Methodology will be useful for unpredictable problem
domains, where all the possible system's situations cannot be considered
beforehand. It could also be useful for creative tasks, where a self-organizing system can explore the design space in an alternative way.

\item Most methodologies in the literature apply to software systems, e.g. 
\cite{JacobsonEtAl1999,Jennings2000}. This one is more general, since it is
domain independent. The principles presented are such that can be applied to any domain for developing a functioning self-organizing system: Any system can be modeled as a group of agents, with satisfactions depending on their goals. The question is \emph{when} is it useful to use this Methodology. Only application of the Methodology will provide an answer to this question. It should be noted that several approaches have been proposed in parallel, e.g. \cite{CaperaEtAl2003,DeWolfHolvoet2005}, that, as the present work, and are part of the ongoing effort by the research community to understand self-organizing systems.

\item The proposed Methodology is not quite a spiral model \cite{Boehm1988},
because the last stage does not need to be reached to return to the first. This is, there is no need to deploy a working version (finish a cycle/iteration) before revisiting a previous stage, as for example in extreme programming. 
Rather, the Methodology is a \textquotedblleft \textit{backtracking} model%
\textquotedblright , where the designer can always return to previous
stages.

\item It is not necessary to understand a solution before testing it. In
many cases understanding can come only after testing, i.e., the global
behavior of the system is irreducible. Certainly, understanding the causes
of a phenomenon will allow the modelers to have a greater control over it.
\end{itemize}

\begin{figure*}[tp]
\begin{center}
\includegraphics[
width= 10.1067cm, height=13.4565cm]
{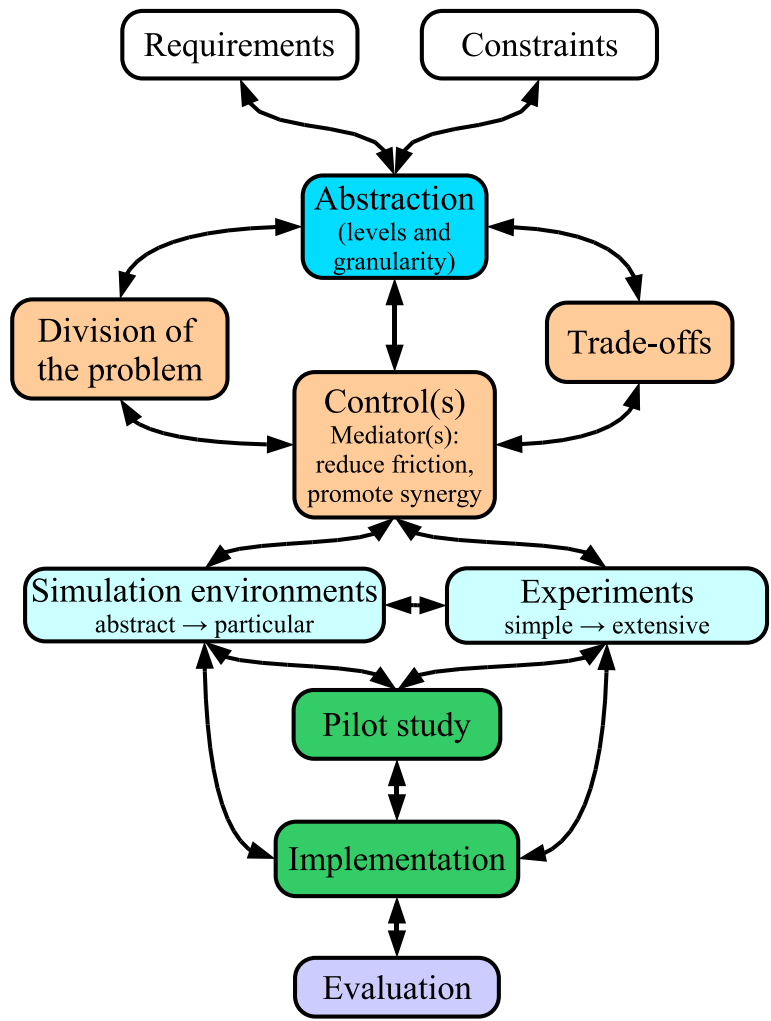}
\end{center}
\caption{Detailed diagram of Methodology}
\label{diagram-detailed}
\end{figure*}

A detailed diagram of the different substeps of the Methodology can be
appreciated in Figure \ref{diagram-detailed}.

\section{Case Study: Self-organizing Traffic Lights}\label{secSOTL}

Recent work on self-organizing traffic lights \cite{Gershenson2005} will be
used to illustrate the flow through the different steps of the Methodology.
These traffic lights are called self-organizing because the global
performance is given by the local rules followed by each traffic light: they
are unaware of the state of other intersections and still manage to achieve
global coordination.

Traffic modeling has increased the understanding of this complex phenomenon 
\cite%
{PrigogineHerman1971,Traffic95,Traffic97,Traffic99,Helbing1997,HelbingHuberman1998}%
. Even when vehicles can follow simple rules, their local interactions
generate global patterns that cannot be reduced to individual behaviors.
Controlling traffic lights in a city is not an easy task: it requires the
coordination of a multitude of components; the components affect one
another; furthermore, these components do not operate at the same pace over
time. Traffic flows and densities change constantly. Therefore, this problem
is suitable to be tackled by self-organization. A centralized system could also perform
the task, but in practice the amount of computation required to process all
the data from a city is too great to be able to respond in real time. Thus, a self-organizing system seems to be a promising alternative.

\textbf{Requirements}. The goal is to develop a feasible and efficient
traffic light control system.

\textbf{Representation}. The traffic light system can be modelled on two
levels: the vehicle level and the city level. These are easy to identify
because vehicles are objects that move through the city, establishing clear
spatiotemporal distinctions. The goal of the vehicles is to flow as fast as
possible, so their ``satisfaction" $\sigma $ can be measured in terms of
their average speed and average waiting time at a red light. Cars will have
a maximum $\sigma $ if they go as fast as they are allowed, and do not stop
at intersections. $\sigma $ would be zero if a car stops indefinitely. The
goal of the traffic light system on the city level is to enable vehicles to
flow as fast as possible, while mediating their conflicts for space and time
at intersections. This would minimize fuel consumption, noise, pollution,
and stress in the population. The satisfaction of the city can be measured
in terms of the average speeds and average waiting times of all vehicles
(i.e. average of $\sigma _{i},\ \forall i$), and with the average percentage
of stationary cars. $\sigma _{sys}$ will be maximum if all cars go as fast
as possible, and are able to flow through the city without stopping. If a
traffic jam occurs and all the vehicles stop, then $\sigma _{sys}$ would be
minimal. 

\textbf{Modeling}. Now the problem for the Control can be formulated: find
a mechanism that will coordinate traffic lights so that these will mediate
between vehicles to reduce their friction (i.e. try to prevent them from
arriving at the same time at crossings). This will maximize the
satisfactions of the vehicles and of the city ($\sigma _{i}$'s and $\sigma
_{sys}$). Since all vehicles contribute equally to $\sigma _{sys}$, ideally
the Control should minimize frictions via Compromise.

\textbf{Simulation}. A simple simulation was developed in NetLogo \cite%
{Wilensky1999}, a multi-agent modeling environment. The ``Gridlock" model 
\cite{WilenskyStroup2002} was extended to implement different traffic
control strategies. It consists of an abstract traffic grid with
intersections between cyclic single-lane arteries of two types: vertical or
horizontal (similar to the scenarios of \cite{BML1992,BrockfeldEtAl2001}).\ Cars only
flow in a straight line, either eastbound or southbound. Each crossroad has
traffic lights that allow traffic flow in only one of the intersecting
arteries with a green light. Yellow or red lights stop the traffic. The
light sequence for a given artery is green-yellow-red-green. Cars simply try
to go at a maximum speed of 1 ``patch" per timestep, but stop when a car or a
red or yellow light is in front of them. Time is discrete, but not space. A
``patch" is a square of the environment the size of a car. The simulation can
be tested at the URL http://homepages.vub.ac.be/\symbol{126}%
cgershen/sos/SOTL/SOTL.html . At first, a tentative model was implemented.
The idea was unsuccessful. However, after refining the model, an efficient
method was discovered, named \emph{sotl-request}.

\textbf{Modeling}. In the \emph{sotl-request} method, each traffic light
keeps a count $\kappa _{i}$ of the number of cars times time steps ($c\ast
ts $) approaching \emph{only} the red light, independently of the status or
speed of the cars (i.e. moving or stopped). $\kappa _{i}$ can be seen as the
integral of waiting/approaching cars over time. When $\kappa _{i}$ reaches a
threshold $\theta $, the opposing green light turns yellow, and the
following time step it turns red with $\kappa _{i}=0$ , while the red light
which counted turns green. In this way, if there are more cars approaching
or waiting before a red light, the light will turn green faster than if
there are only few cars. This simple mechanism achieves self-organization as
follows: if there is a single or just a few cars, these will be made to stop
for a longer period before a red light. This gives time for other cars to
join them. As more cars join the group, cars will be made to wait shorter
periods before a red light. Once there are enough cars, the red light will
turn green even before the oncoming cars reach the intersection, thereby
generating ``green waves". Having ``platoons" or ``convoys" of cars moving
together improves traffic flow, compared to a homogeneous distribution of
cars, since there are large empty areas between platoons, which can be used
by crossing platoons with few interferences. The \emph{sotl-request} method
has no phase or internal clock. Traffic lights change only when the above
conditions are met. If no cars are approaching a red light, the
complementary light can remain green.

\textbf{Representation}. It becomes clear now that it would be useful to
consider traffic lights as agents as well. Their goal is to ``get rid" of
cars as quickly as possible. To do so, they should avoid having green lights
on empty streets and red lights on streets with high traffic density. Since
the satisfactions of the traffic lights and vehicles are complementary, they
should interact via Cooperation to achieve synergy. Also, $\sigma _{sys}$
could be formulated in terms of the satisfactions of traffic lights,
vehicles, or both.

\textbf{Modeling}. Two classic methods were implemented to compare their
performance with \emph{sotl-request}: \emph{marching} and \emph{optim}.

\emph{Marching} is a very simple method. All traffic lights ``march in step":
all green lights are either southbound or eastbound, synchronized in time.
Intersections have a phase $\varphi _{i}$, which counts time steps. $\varphi
_{i}$ is reset to zero when the phase reaches a period value $p$. When $%
\varphi _{i}==0$, red lights turn green, and yellow lights turn red. Green
lights turn yellow one time step earlier, i.e. when $\varphi ==p-1$. A full
cycle of an intersection consists of $2p$ time steps. ``Marching"
intersections are such that $\varphi _{i}==\varphi _{j},\forall i,j$.

The \emph{optim} method is implemented trying to set phases $\varphi _{i}$
of traffic lights so that, as soon as a red light turns green, a car that
was made to stop would find the following traffic lights green. In other
words, a fixed solution is obtained so that \emph{green waves} flow to the
southeast. The simulation environment has a radius of $r$ square patches, so
that these can be identified with coordinates $(x_{i},y_{i}),$ $%
x_{i},y_{i}\in \lbrack -r,r]$. Therefore, each artery consists of $2r+1$
patches. In order to synchronize all the intersections, red lights should
turn green and yellow lights should turn red when

\begin{equation}
\varphi _{i}==round(\frac{2r+x_{i}-y_{i}}{4})
\end{equation}

and green lights should turn to yellow the previous time step. The period
should be $p=r+3$. The three is added as an extra margin for the reaction
and acceleration times of cars (found to be best, for low densities, by
trial and error).

These two methods are \emph{non-adaptive}, in the sense that their behavior
is dictated beforehand, and they do not consider the actual state of the
traffic. Therefore, there cannot be Cooperation between vehicles and traffic
lights, since the latter ones have fixed behaviors. On the other hand,
traffic lights under the \emph{sotl-request} method are sensitive to the
current traffic condition, and can therefore respond to the needs of the
incoming vehicles.

\textbf{Simulation}. Preliminary experiments have shown that \emph{%
sotl-request}, compared with the two traditional methods, achieves very good
results for low traffic densities, but very poor results for high traffic
densities. This is because depending on the value of $\theta $, high traffic
densities can cause the traffic lights to change too fast. This obstructs
traffic flow. A new model was developed, taking this factor into account.

\textbf{Modeling}. The \emph{sotl-phase} method takes \emph{sotl-request}
and only adds the following constraint: a traffic light will not be changed
if the time passed since the last light change is less than a minimum phase,
i.e. $\varphi _{i}<\varphi _{\min }$. Once $\varphi _{i}\geq \varphi _{\min
} $, the lights will change if/when $\kappa _{i}$ $\geq $ $\theta $. This
prevents the fast changing of lights\footnote{%
A similar method has been used successfully in the United Kingdom for some
time, but for isolated intersections \cite{VincentYoung1986}.}.

\textbf{Simulation}. \emph{Sotl-phase} performed a bit less effectively than 
\emph{sotl-request} for very low traffic densities, but still much better
than the classic methods. However, \emph{sotl-phase} outperformed them also
for high densities. An unexpected phenomenon was observed: for certain
traffic densities, \emph{sotl-phase} achieved \emph{full synchronization},
in the sense that no car stopped. Therefore, speeds were maximal and there
were no waiting times nor sopped cars. Thus, satisfaction was maximal for
vehicles, traffic lights, and the city. Still, this is not a realistic
situation, because full synchronization is achieved due to the toroidal
topology of the simulation environment. The full synchronization is achieved
because platoons are promoted by the traffic lights, and platoons can move
faster through the city modulating traffic lights. If two platoons are
approaching an intersection, \emph{sotl-phase} will stop one of them, and
allow the other to pass without stopping. The latter platoon keeps its phase
as it goes around the torus, and the former adjusts its speed until it finds
a phase that does not cause a conflict with another platoon.

\textbf{Modeling}. Understanding the behavior of the platoons, it can be
seen that there is a favorable system/context trade-off. There is no need
of direct communication between traffic lights, since information can
actually be sent via platoons of vehicles. The traffic lights communicate 
\emph{stigmergically} \cite{TheraulazBonabeau1999}, i.e. via their
environment, where the vehicles are conceptualized as the environment of
traffic lights.

\textbf{Simulation}. With encouraging results, changes were made to the
Simulation to make it more realistic. Thus, a scenario similar to the one of 
\cite{FaietaHuberman1993} was developed. Traffic flow in four directions was
introduced, alternating streets. This is, arteries still consist of one
lane, but the directions alternate: southbound-northbound in vertical roads,
and eastbound-westbound in horizontal roads. Also, the possibility of having
more cars flowing in particular directions was introduced. Peak hour traffic
can be simulated like this, regulating the percentages of cars that will
flow in different roads. An option to switch off the torus in the simulation
was added. Finally, a probability of turning at an intersection \thinspace $
P_{turn}$ was included. Therefore, when a car reaches an intersection, it
will have a probability \thinspace $P_{turn}$ of reducing its speed and
turning in the direction of the crossing street. This can cause cars to
leave platoons, which are more stable when $P_{turn}=0$.

The results of experiments in the more realistic Simulation confirmed the
previous ones: self-organizing methods outperform classic ones. There can
still be full synchronization with alternating streets, but not without a
torus or with $P_{turn}>0$.

\textbf{Modeling}. Another method was developed, \emph{sotl-platoon}, adding two restrictions to \emph{sotl-phase} for
regulating the size of platoons. Before changing a red light to green, \emph{
sotl-platoon} checks if a platoon is not crossing through, not to break it.
More precisely, a red light is not changed to green if on the crossing
street there is at least one car approaching at $\omega $ patches from the
intersection. This keeps platoons together. For high densities, this
restriction alone would cause havoc, since large platoons would block the
traffic flow of intersecting streets. To avoid this, a second restriction is
introduced. The first restriction is not taken into account if there are
more than $\mu $ cars approaching the intersection. Like this, long platoons
can be broken, and the restriction only comes into place if a platoon will
soon be through an intersection.

\textbf{Simulation}. \emph{Sotl-platoon} manages to keep platoons together,
achieving full synchronization commonly for a wide density range, more
effectively than \emph{sotl-phase} (when the torus is active). This is
because the restrictions of \emph{sotl-platoon} prevent the breaking of
platoons when these would leave few cars behind, with a small time cost for
waiting vehicles. Still, this cost is much lower than breaking a platoon and
waiting for separated vehicles to join back again so that they can switch
red lights to green before reaching an intersection. However, for high
traffic densities platoons aggregate too much, making traffic jams more
probable. The \emph{sotl-platoon} method fails when a platoon waiting to
cross a street is long enough to reach the previous intersection, but not
long enough to cut its tail. This will prevent waiting cars from advancing,
until more cars join the long platoon. This failure could probably be
avoided introducing further restrictions. In more realistic experiments
(four directions, no torus, $P_{turn}=0.1$), \emph{sotl-platoon} gives on
average 30\% (up to 40\%) more average speed, half the stopped cars, and
seven times less average waiting times than non-responsive methods. Complete
results and graphics of the experiments discussed here can be found in \cite
{Gershenson2005}.

\textbf{Representation}. If priority is to be given to certain vehicles
(e.g. public transport, emergency), weights can be added to give more
importance to some $\sigma _{i}$'s.

A meso-level might be considered, where properties of platoons can be
observed: their behaviors, performance, and satisfaction and the
relationships of these with the vehicle and city levels could enhance the
understanding of the self-organizing traffic lights and even improve them.

\textbf{Simulation}. Streets of varying distances between crossings were
tested, and all the self-organizing methods maintained their good
performance. Still more realistic simulations should be made before moving
to the Implementation, because of the cost of such a system. At least,
multiple-street intersections, multiple-lane streets, lane changing,
different driving behaviors, and non homogeneous streets should be
considered.

\textbf{Application}. The proposed system has not been implemented yet.
Still, it is feasible to do so, since there is the sensor technology to
implement the discussed methods in an affordable way. Currently, a more realistic simulation is being developed in cooperation with the Brussels Ministry of Mobility and Public Works to study its potential application in the city of Brussels.
A pilot study should
be made before applying it widely, to fine tune different parameters and
methods. External factors, e.g. pedestrians and cyclists, could also affect the
performance of the system.

Pedestrians could be taken into account considering them as cars approaching
a red light. For example, a button could be used to inform the intersection
of their presence, and this would contribute to the count $\kappa _{i}$.

A mixed strategy between different methods could be considered, e.g. \emph{%
sotl-platoon} for low and medium densities, and \emph{sotl-phase} or \emph{%
marching} for high densities.

\textbf{Evaluation}. If a city deploys a self-organizing traffic light
system, it should be monitored and compared with previous systems. This will
help to improve the system. If the system would be an affordable success, its implementation in other cities would be promoted.

\section{Discussion}\label{secDiscussion}

As could be seen in the case study, the backtracking between different steps
in the Methodology is necessary because the behavior of the system cannot
be predicted from the Modeling, i.e. it is not reducible. It might be
possible to reason about all possible outcomes of simple systems, and then
to implement the solution. But when complexity needs to be dealt with, a
mutual feedback between experience and reasoning needs to be established,
since reasoning alone cannot process all the information required to predict
the behavior of a complex system \cite{Edmonds2005}.

For this same reason, it would be preferable for the Control to be
distributed. Even when a central supercomputer could possibly solve a
problem in real time, the information delay caused by data transmission and
integration can reduce the efficiency of the system. Also, a distributed
Control will be more robust, in as much as if a module malfunctions, the
rest of the system can still provide reliable solutions. If a central
Control fails, the whole system will stop working.

The Simulation and Experiments are strictly necessary in the design of self-organizing systems \cite{Edmonds2005}. This is because their performance cannot be evaluated by purely formal methods \cite{EdmondsBryson2004}. Still, formal methods are necessary in the first stages of the Methodology. I am not suggesting a trial-and-error search. But since the behavior of a complex system in a complex environment cannot be predicted completely, the models need to be contrasted with experimentation before they can be validated.
This Methodology suggests one possible path for finding solutions.

Now the reader might wonder whether the proposed Methodology is a \emph{%
top-down} or a \emph{bottom-up} approach. And the answer is: it is both and
neither, since (at least) higher and lower levels of abstraction need to be
considered simultaneously. The approach tests different local behaviors,
and observes local and global (and meso) performances, for local and global
(and meso) requirements. Thus, the Methodology can be seen as a \emph{multi-level} approach.

Since ``conflicts" between agents need to be solved at more than one level,
the Control strategies should be carefully chosen and tested. A situation as
in the prisoner's dilemma \cite{Axelrod1984} might easily arise, when the
``best" solution on one level/timescale is not the best solution on another
level/timescale.

Many frictions between agents are due to faulty communication, especially in
social and political relations. If agents do not \textquotedblleft
know\textquotedblright\ the goals of others, it will be much more difficult
to coordinate and increase $\sigma _{sys}$. For example, in a social system,
knowing what people or corporations need to fulfill their goals is not so
obvious. Still, with emerging technologies, social systems perform better in
this respect. Already in the early 1970s, the project Cybersin in Chile
followed this path \cite{MillerMedina2005}: it kept a daily log of
productions and requirements from all over the country (e.g. mines,
factories, etc.), in order to distribute products where they were needed
most; and as quickly as possible. Another step towards providing faster
response to the needs of both individuals and social systems can be found in
e-government \cite{LayneLee2001}. A company should also follow these
principles to be able to adapt as quickly as possible. It needs to develop
"sensors" to perceive the satisfactions and conflicts of agents at different
levels of abstraction, and needs to develop fast ways of adapting to
emerging conflicts, as well as to changing economic environment. A tempting
solution might be to develop a homogeneous system since, e.g., homogeneous
societies have fewer conflicts \cite{Durkheim1893}. This is because all the
elements of a homogeneous system pursue the same goals. Thus, less diversity
is easier to control. However, less diversity will be less able to adapt to
sudden changes. Nevertheless, societies cannot be \emph{made} homogeneous
without generating conflicts since people are already diverse, and therefore
already have a diversity of goals. The legacy \cite{ValckenaersEtAl2003}\ of
social systems gives less freedom to a designer, since some goals are
already within the system. A social Control/mediator needs to satisfy these
while trying to satisfy those of the social system.

\section{Conclusions}\label{secConclusions}

This paper suggests a conceptual framework and a general methodology for
designing and controlling self-organizing systems. The Methodology proposes
the exploration for proper Control mechanisms/mediators/constraints that
will reduce frictions and promote synergy so that elements will dynamically
reach a robust and efficient solution. The proposed Methodology is general, but certainly it is not the only way to \emph{describe} self-organizing systems.

Even if this paper is aimed mainly at engineers, it is rather philosophical.
It presents no concrete results, but \emph{ideas} that can be exploited to
produce them. Certainly, these ideas have their roots in current practices, and many of them are not novel. Still, the aim of this work is not for novelty but for synthesis.

The Methodology strives to build artificial systems. Still, these could be used to understand natural systems using the synthetic method \cite{Steels1993}. Therefore, the ideas presented here are potentially useful not only for engineering, but also for science.

The backtracking ideology is also applicable to this Methodology: it will be
improved once applied, through learning from experience. This Methodology is not final, but evolving. The more this
Methodology is used, and in a wider variety of areas, the more potentially useful its abstractions will be.
For example, would it be a good strategy to minimize the standard deviation
of $\sigma $'s? This might possibly increase stability and reduce the
probability of conflict, but this strategy, as any other, needs to be tested before it can be properly understood. It is worth noting that apart from self-organizing traffic lights \cite{Gershenson2005}, the Methodology is currently being used to develop Ambient Intelligence protocols \cite{GershensonHeylighen2004} and to study self-organizing bureaucracies.

Any system is liable to make mistakes (and \emph{will} make them in an
unpredictable environment). But a good system will \emph{learn} from its
mistakes. This is the basis for adaptation. It is pointless to attempt to
build a ``perfect" system, since it is not possible to predict future
interactions with its environment. What should be done is to build systems
that can adapt to their unexpected future and are robust enough not to be destroyed in the attempt. Self-organization provides one way to achieve this, but there is still much to be done to harness its full potential.

\begin{acks}
I should like to thank Hugues Bersini, Marco Dorigo, Erden G\"{o}ktepe,
Dirk Helbing, Francis Heylighen, Diana Mangalagiu, Peter McBurney, Juan Juli\'{a}n Merelo, Marko Rodriguez, Frank Schweitzer, Sorin Solomon, and Franco Zambonelli for interesting discussions and
comments. I also wish to thank Michael Whitburn for proof-reading an earlier version of the manuscript. This research was partly supported by the Consejo Nacional de
Ciencia y Teconolg\'{\i}a (CONACyT) of M\'{e}xico.
\end{acks}

\small{ 
\bibliographystyle{acmtrans_cgg}
\bibliography{carlos,COG,RBN,sos,traffic,evolution}

\begin{thebibliography}{}

\bibitem[\protect\citeauthoryear{Anderson}{Anderson}{1972}]{Anderson1972}
{\sc Anderson, P.~W.} 1972.
\newblock More is different.
\newblock {\em Science\/}~{\em 177}, 393--396.

\bibitem[\protect\citeauthoryear{Ashby}{Ashby}{1947}]{Ashby1947}
{\sc Ashby, W.~R.} 1947.
\newblock The nervous system as physical machine: With special reference to the
  origin of adaptive behavior.
\newblock {\em Mind\/}~{\em 56,\/}~221 (January), 44--59.

\bibitem[\protect\citeauthoryear{Ashby}{Ashby}{1956}]{Ashby1956}
{\sc Ashby, W.~R.} 1956.
\newblock {\em An Introduction to Cybernetics}.
\newblock Chapman \& Hall, London.
\newblock http://pcp.vub.ac.be/ASHBBOOK.html.

\bibitem[\protect\citeauthoryear{Ashby}{Ashby}{1962}]{Ashby1962}
{\sc Ashby, W.~R.} 1962.
\newblock Principles of the self-organizing system.
\newblock In {\em Principles of Self-Organization}, {H.~V. Foerster} {and}
  {J.~G.~W.~Zopf}, Eds. Pergamon, 255--278.

\bibitem[\protect\citeauthoryear{Axelrod}{Axelrod}{1984}]{Axelrod1984}
{\sc Axelrod, R.~M.} 1984.
\newblock {\em The Evolution of Cooperation}.
\newblock Basic Books, New York.

\bibitem[\protect\citeauthoryear{Bar-Yam}{Bar-Yam}{1997}]{Bar-Yam1997}
{\sc Bar-Yam, Y.} 1997.
\newblock {\em Dynamics of Complex Systems}.
\newblock Studies in Nonlinearity. Westview Press.
\newblock http://www.necsi.org/publications/dcs/.

\bibitem[\protect\citeauthoryear{Bar-Yam}{Bar-Yam}{2005}]{Bar-Yam2005}
{\sc Bar-Yam, Y.} 2005.
\newblock About engineering complex systems: Multiscale analysis and
  evolutionary engineering.
\newblock See \citeN{Edmonds2005}, 16--31.
\newblock http://necsi.org/projects/yaneer/ESOA04.pdf.

\bibitem[\protect\citeauthoryear{Beer}{Beer}{1966}]{Beer1966}
{\sc Beer, S.} 1966.
\newblock {\em Decision and Control}.
\newblock John Wiley and Sons, New York.

\bibitem[\protect\citeauthoryear{Berners-Lee, Hendler, and Lassila}{Berners-Lee
  et~al\mbox{.}}{2001}]{Berners-LeeEtAl2001}
{\sc Berners-Lee, T.}, {\sc Hendler, J.}, {\sc and} {\sc Lassila, O.} 2001.
\newblock The semantic web link to this article e-mail this article
  printer-friendly version subscribe the semantic web link to this article
  e-mail this article printer-friendly version subscribe the semantic web.
\newblock {\em Scientific American\/}.
\newblock
  http://www.sciam.com/article.cfm?articleID=00048144-10D2-1C70-84A9809EC588EF%
21.

\bibitem[\protect\citeauthoryear{Biham, Middleton, and Levine}{Biham
  et~al\mbox{.}}{1992}]{BML1992}
{\sc Biham, O.}, {\sc Middleton, A.~A.}, {\sc and} {\sc Levine, D.} 1992.
\newblock Self-organization and a dynamical transition in traffic-flow models.
\newblock {\em Physical Review A\/}~{\em 46}, R6124--R6127.
\newblock http://dx.doi.org/10.1103/PhysRevA.46.R6124.

\bibitem[\protect\citeauthoryear{Boehm}{Boehm}{1988}]{Boehm1988}
{\sc Boehm, B.~W.} 1988.
\newblock A spiral model of software development and enhancement.
\newblock {\em Computer\/}~{\em 21,\/}~5, 61--72.
\newblock http://dx.doi.org/10.1109/2.59.

\bibitem[\protect\citeauthoryear{Bonabeau, Dorigo, and Theraulaz}{Bonabeau
  et~al\mbox{.}}{1999}]{BonabeauEtAl1999}
{\sc Bonabeau, E.}, {\sc Dorigo, M.}, {\sc and} {\sc Theraulaz, G.} 1999.
\newblock {\em Swarm Intelligence: From Natural to Artificial Systems}.
\newblock Santa Fe Institute Studies in the Sciences of Complexity. Oxford
  University Press, New York.

\bibitem[\protect\citeauthoryear{Brockfeld, Barlovic, Schadschneider, and
  Schreckenberg}{Brockfeld et~al\mbox{.}}{2001}]{BrockfeldEtAl2001}
{\sc Brockfeld, E.}, {\sc Barlovic, R.}, {\sc Schadschneider, A.}, {\sc and}
  {\sc Schreckenberg, M.} 2001.
\newblock Optimizing traffic lights in a cellular automaton model for city
  traffic.
\newblock {\em Physical Review E\/}~{\em 64}, 056132.
\newblock http://dx.doi.org/10.1103/PhysRevE.64.056132.

\bibitem[\protect\citeauthoryear{Camazine, Deneubourg, Franks, Sneyd,
  Theraulaz, and Bonabeau}{Camazine et~al\mbox{.}}{2003}]{CamazineEtAl2003}
{\sc Camazine, S.}, {\sc Deneubourg, J.-L.}, {\sc Franks, N.~R.}, {\sc Sneyd,
  J.}, {\sc Theraulaz, G.}, {\sc and} {\sc Bonabeau, E.} 2003.
\newblock {\em Self-Organization in Biological Systems}.
\newblock Princeton University Press.
\newblock http://www.pupress.princeton.edu/titles/7104.html.

\bibitem[\protect\citeauthoryear{Capera, Georg{\'e}, Gleizes, and Glize}{Capera
  et~al\mbox{.}}{2003}]{CaperaEtAl2003}
{\sc Capera, D.}, {\sc Georg{\'e}, J.-P.}, {\sc Gleizes, M.-P.}, {\sc and} {\sc
  Glize, P.} 2003.
\newblock The {AMAS} theory for complex problem solving based on
  self-organizing cooperative agents.
\newblock In {\em 1st International Workshop on Theory and Practice of Open
  Computational Systems {TAPOCS} 2003 at {IEEE} 12th International Workshop on
  Enabling Technologies: Infrastructure for Collaborative Enterprises {WETICE}
  2003}. 383.
\newblock
  http://csdl.computer.org/comp/proceedings/wetice/2003/1963/00/19630383abs.ht%
m.

\bibitem[\protect\citeauthoryear{Corning}{Corning}{2003}]{Corning2003}
{\sc Corning, P.~A.} 2003.
\newblock {\em Nature's Magic: Synergy in Evolution and the Fate of Humankind}.
\newblock Cambridge University Press.
\newblock http://www.complexsystems.org/magic.html.

\bibitem[\protect\citeauthoryear{Cotton}{Cotton}{1996}]{Cotton1996}
{\sc Cotton, T.} 1996.
\newblock Evolutionary fusion: A customer-oriented incremental life-cycle for
  {Fusion}.
\newblock {\em Hewlett Packard Journal\/}~{\em 47,\/}~4.
\newblock http://www.hpl.hp.com/hpjournal/96aug/aug96a3.htm.

\bibitem[\protect\citeauthoryear{de~Jong}{de~Jong}{2000}]{DeJong2000}
{\sc de~Jong, E.~D.} 2000.
\newblock Autonomous formation of concepts and communication.
\newblock Ph.D. thesis, Vrije Universiteit Brussel.
\newblock http://www.cs.uu.nl/\%7Edejong/thesis/.

\bibitem[\protect\citeauthoryear{{De Wolf} and Holvoet}{{De Wolf} and
  Holvoet}{2005}]{DeWolfHolvoet2005}
{\sc {De Wolf}, T.} {\sc and} {\sc Holvoet, T.} 2005.
\newblock Towards a methodolgy for engineering self-organising emergent
  systems.
\newblock In {\em Self-Organization and Autonomic Informatics (I)}, {H.~Czap},
  {R.~Unland}, {C.~Branki}, {and} {H.~Tianfield}, Eds. Frontiers in Artificial
  Intelligence and Applications, vol. 135. IOS Press, 18--34.
\newblock http://www.cs.kuleuven.ac.be/~tomdw/publications/pdfs/2005soas.pdf.

\bibitem[\protect\citeauthoryear{{De Wolf}, Samaey, and Holvoet}{{De Wolf}
  et~al\mbox{.}}{2005}]{DeWolfEtAl2005}
{\sc {De Wolf}, T.}, {\sc Samaey, G.}, {\sc and} {\sc Holvoet, T.} 2005.
\newblock Engineering self-organising emergent systems with simulation-based
  scientific analysis.
\newblock In {\em Proceedings of the International Workshop on Engineering
  Self-Organising Applications}. Utrecht, The Netherlands,.
\newblock
  http://www.cs.kuleuven.ac.be/~tomdw/publications/pdfs/2005esoa05proc.pdf.

\bibitem[\protect\citeauthoryear{{Di Marzo Serugendo}}{{Di Marzo
  Serugendo}}{2004}]{DiMarzoSerugendo2004}
{\sc {Di Marzo Serugendo}, G.} 2004.
\newblock Trust as an interaction mechanism for self-organising systems.
\newblock In {\em International Conference on Complex Systems {(ICCS'04)}},
  {Y.~Bar-Yam}, Ed.

\bibitem[\protect\citeauthoryear{{Di Marzo Serugendo}, Karageorgos, Rana, and
  Zambonelli}{{Di Marzo Serugendo} et~al\mbox{.}}{2004}]{EngineeringSOS2004}
{\sc {Di Marzo Serugendo}, G.}, {\sc Karageorgos, A.}, {\sc Rana, O.~F.}, {\sc
  and} {\sc Zambonelli, F.}, Eds. 2004.
\newblock {\em Engineering Self-Organising Systems, Nature-Inspired Approaches
  to Software Engineering}. Lecture Notes in Computer Science, vol. 2977.
  Springer.
\newblock Revised and extended papers presented at the Engineering
  Self-Organising Applications Workshop, ESOA 2003, held at AAMAS 2003 in
  Melbourne, Australia, in July 2003 and selected invited papers from leading
  researchers in self-organisation.

\bibitem[\protect\citeauthoryear{Dorigo, Trianni, \c{S}ahin, Gro{\ss}, Labella,
  Baldassarre, Nolfi, Deneubourg, Mondada, Floreano, and Gambardella}{Dorigo
  et~al\mbox{.}}{2004}]{DorigoEtAl2004}
{\sc Dorigo, M.}, {\sc Trianni, V.}, {\sc \c{S}ahin, E.}, {\sc Gro{\ss}, R.},
  {\sc Labella, T.~H.}, {\sc Baldassarre, G.}, {\sc Nolfi, S.}, {\sc
  Deneubourg, J.-L.}, {\sc Mondada, F.}, {\sc Floreano, D.}, {\sc and} {\sc
  Gambardella, L.} 2004.
\newblock Evolving self-organizing behaviors for a swarm-bot.
\newblock {\em Autonomous Robots\/}~{\em 17,\/}~2-3, 223--245.
\newblock http://www.swarm-bots.org.

\bibitem[\protect\citeauthoryear{Durkheim}{Durkheim}{1984}]{Durkheim1893}
{\sc Durkheim, {\'E}.} 1893 (1984).
\newblock {\em The Division of Labor in Society}.
\newblock The Free Press, New York.
\newblock Translated by George Simpson.

\bibitem[\protect\citeauthoryear{Edmonds}{Edmonds}{1999}]{Edmonds1999}
{\sc Edmonds, B.} 1999.
\newblock What is complexity?: the philosophy of complexity per se with
  application to some examples in evolution.
\newblock In {\em The Evolution of Complexity}, {F.~Heylighen}, {J.~Bollen},
  {and} {A.~Riegler}, Eds. Kluwer, Dordrecht, 1--18.
\newblock http://bruce.edmonds.name/evolcomp/.

\bibitem[\protect\citeauthoryear{Edmonds}{Edmonds}{2005}]{Edmonds2005}
{\sc Edmonds, B.} 2005.
\newblock Using the experimental method to produce reliable self-organised
  systems.
\newblock In {\em Engineering Self Organising Sytems: Methodologies and
  Applications}, {S.~Brueckner}, {G.~Serugendo-Di~Marzo}, {A.~Karageorgos},
  {and} {R.~Nagpal}, Eds. Lecture Notes in Artificial Intelligence, vol. 3464.
  Springer, 84--99.
\newblock http://cfpm.org/cpmrep131.html.

\bibitem[\protect\citeauthoryear{Edmonds and Bryson}{Edmonds and
  Bryson}{2004}]{EdmondsBryson2004}
{\sc Edmonds, B.} {\sc and} {\sc Bryson, J.} 2004.
\newblock The insufficiency of formal design methods - the necessity of an
  experimental approach for the understanding and control of complex mas.
\newblock In {\em Proceedings of the 3rd International Joint Conference on
  Autonomous Agents \& Multi Agent Systems (AAMAS'04)}, {N.~Jennings},
  {C.~Sierra}, {L.~Sonenberg}, {and} {M.~Tambe}, Eds. ACM Press, New York,
  938--945.
\newblock http://cfpm.org/cpmrep128.html.

\bibitem[\protect\citeauthoryear{Faieta and Huberman}{Faieta and
  Huberman}{1993}]{FaietaHuberman1993}
{\sc Faieta, B.} {\sc and} {\sc Huberman, B.~A.} 1993.
\newblock Firefly: A synchronization strategy for urban traffic control.
\newblock Tech. Rep. SSL-42, Xerox PARC, Palo Alto.

\bibitem[\protect\citeauthoryear{Fern{\'a}ndez and Sol{\'e}}{Fern{\'a}ndez and
  Sol{\'e}}{2004}]{FernandezSole2003}
{\sc Fern{\'a}ndez, P.} {\sc and} {\sc Sol{\'e}, R.} 2004.
\newblock The role of computation in complex regulatory networks.
\newblock In {\em Power Laws, Scale-Free Networks and Genome Biology}, {E.~V.
  Koonin}, {Y.~I. Wolf}, {and} {G.~P. Karev}, Eds. Landes Bioscience.
\newblock http://arxiv.org/abs/q-bio.MN/0311012.

\bibitem[\protect\citeauthoryear{Gaines}{Gaines}{1994}]{Gaines1994}
{\sc Gaines, B.~R.} 1994.
\newblock The collective stance in modeling expertise in individuals and
  organizations.
\newblock {\em Int. J. Expert Systems\/}~{\em 71}, 22--51.

\bibitem[\protect\citeauthoryear{G{\"a}rdenfors}{G{\"a}rdenfors}{2000}]{Garden%
fors2000}
{\sc G{\"a}rdenfors, P.} 2000.
\newblock {\em Conceptual Spaces: The Geometry of Thought}.
\newblock Bradford Books. MIT Press.
\newblock http://mitpress.mit.edu/0262572192.

\bibitem[\protect\citeauthoryear{Gershenson}{Gershenson}{2002}]{Gershenson2002%
a}
{\sc Gershenson, C.} 2002.
\newblock Complex philosophy.
\newblock In {\em Proceedings of the 1st Biennial Seminar on Philosophical,
  Methodological $\And$ Epistemological Implications of Complexity Theory}. La
  Habana, Cuba.
\newblock http://arXiv.org/abs/nlin.AO/0109001.

\bibitem[\protect\citeauthoryear{Gershenson}{Gershenson}{2004a}]{Gershenson200%
4}
{\sc Gershenson, C.} 2004a.
\newblock Cognitive paradigms: Which one is the best?
\newblock {\em Cognitive Systems Research\/}~{\em 5,\/}~2 (June), 135--156.
\newblock http://dx.doi.org/10.1016/j.cogsys.2003.10.002.

\bibitem[\protect\citeauthoryear{Gershenson}{Gershenson}{2004b}]{Gershenson200%
4c}
{\sc Gershenson, C.} 2004b.
\newblock Introduction to random boolean networks.
\newblock In {\em Workshop and Tutorial Proceedings, Ninth International
  Conference on the Simulation and Synthesis of Living Systems {(ALife} {IX)}},
  {M.~Bedau}, {P.~Husbands}, {T.~Hutton}, {S.~Kumar}, {and} {H.~Suzuki}, Eds.
  Boston, MA, 160--173.
\newblock http://uk.arxiv.org/abs/nlin.AO/0408006.

\bibitem[\protect\citeauthoryear{Gershenson}{Gershenson}{2005}]{Gershenson2005}
{\sc Gershenson, C.} 2005.
\newblock Self-organizing traffic lights.
\newblock {\em Complex Systems\/}~{\em 16,\/}~1, 29--53.
\newblock http://uk.arxiv.org/abs/nlin.AO/0411066.

\bibitem[\protect\citeauthoryear{Gershenson, Broekaert, and Aerts}{Gershenson
  et~al\mbox{.}}{2003}]{GershensonEtAl2003a}
{\sc Gershenson, C.}, {\sc Broekaert, J.}, {\sc and} {\sc Aerts, D.} 2003.
\newblock Contextual random {Boolean} networks.
\newblock In {\em Advances in Artificial Life, 7th European Conference, {ECAL}
  2003 {LNAI} 2801}, {W.~Banzhaf}, {T.~Christaller}, {P.~Dittrich}, {J.~T.
  Kim}, {and} {J.~Ziegler}, Eds. Springer-Verlag, 615--624.
\newblock http://arxiv.org/abs/nlin.AO/0303021.

\bibitem[\protect\citeauthoryear{Gershenson and Heylighen}{Gershenson and
  Heylighen}{2003}]{GershensonHeylighen2003a}
{\sc Gershenson, C.} {\sc and} {\sc Heylighen, F.} 2003.
\newblock When can we call a system self-organizing?
\newblock In {\em Advances in Artificial Life, 7th European Conference, {ECAL}
  2003 {LNAI} 2801}, {W.~Banzhaf}, {T.~Christaller}, {P.~Dittrich}, {J.~T.
  Kim}, {and} {J.~Ziegler}, Eds. Springer-Verlag, 606--614.
\newblock http://arxiv.org/abs/nlin.AO/0303020.

\bibitem[\protect\citeauthoryear{Gershenson and Heylighen}{Gershenson and
  Heylighen}{2004}]{GershensonHeylighen2004}
{\sc Gershenson, C.} {\sc and} {\sc Heylighen, F.} 2004.
\newblock Protocol requirements for self-organizing artifacts: Towards an
  ambient intelligence.
\newblock In {\em Proceedings of International Conference on Complex Systems
  {ICCS2004}}, {Y.~Bar-Yam}, Ed. Boston, MA.
\newblock Also AI-Lab Memo 04-04, http://uk.arxiv.org/abs/nlin.AO/0404004.

\bibitem[\protect\citeauthoryear{Gershenson and Heylighen}{Gershenson and
  Heylighen}{2005}]{GershensonHeylighen2005}
{\sc Gershenson, C.} {\sc and} {\sc Heylighen, F.} 2005.
\newblock How can we think the complex?
\newblock In {\em Managing Organizational Complexity: Philosophy, Theory and
  Application}, {K.~Richardson}, Ed. Information Age Publishing, Chapter~3.
\newblock http://uk.arxiv.org/abs/nlin.AO/0402023.

\bibitem[\protect\citeauthoryear{Gershenson, Kauffman, and
  Shmulevich}{Gershenson et~al\mbox{.}}{2006}]{GershensonEtAl2006}
{\sc Gershenson, C.}, {\sc Kauffman, S.~A.}, {\sc and} {\sc Shmulevich, I.}
  2006.
\newblock The role of redundancy in the robustness of random boolean networks.
\newblock In {\em Artificial Life X, Proceedings of the Tenth International
  Conference on the Simulation and Synthesis of Living Systems.} MIT Press.
\newblock http://uk.arxiv.org/abs/nlin.AO/0511018.

\bibitem[\protect\citeauthoryear{Haken}{Haken}{1981}]{Haken1981}
{\sc Haken, H.} 1981.
\newblock Synergetics and the problem of selforganization.
\newblock In {\em Self-Organizing Systems: An Interdisciplinary Approach},
  {G.~Roth} {and} {H.~Schwegler}, Eds. Campus Verlag, 9--13.

\bibitem[\protect\citeauthoryear{Hales and Edmonds}{Hales and
  Edmonds}{2003}]{HalesEdmonds2003}
{\sc Hales, D.} {\sc and} {\sc Edmonds, B.} 2003.
\newblock Evolving social rationality for {MAS} using ``tags".
\newblock In {\em Proceedings of the 2nd International Conference on Autonomous
  Agents and Multiagent Systems}, {J.~S. Rosenschein}, {T.~Sandholm},
  {M.~Wooldridge}, {and} {M.~Yokoo}, Eds. ACM Press, 497--503.

\bibitem[\protect\citeauthoryear{Helbing}{Helbing}{1997}]{Helbing1997}
{\sc Helbing, D.} 1997.
\newblock {\em Verkehrsdynamik}.
\newblock Springer, Berlin.

\bibitem[\protect\citeauthoryear{Helbing, Herrmann, Schreckenberg, and
  Wolf}{Helbing et~al\mbox{.}}{2000}]{Traffic99}
{\sc Helbing, D.}, {\sc Herrmann, H.~J.}, {\sc Schreckenberg, M.}, {\sc and}
  {\sc Wolf, D.~E.}, Eds. 2000.
\newblock {\em Traffic and Granular Flow '99: Social, Traffic, and Granular
  Dynamics}. Springer, Berlin.

\bibitem[\protect\citeauthoryear{Helbing and Huberman}{Helbing and
  Huberman}{1998}]{HelbingHuberman1998}
{\sc Helbing, D.} {\sc and} {\sc Huberman, B.~A.} 1998.
\newblock Coherent moving states in highway traffic.
\newblock {\em Nature\/}~{\em 396}, 738--740.

\bibitem[\protect\citeauthoryear{Heylighen}{Heylighen}{2003a}]{Heylighen2003}
{\sc Heylighen, F.} 2003a.
\newblock Mediator evolution.
\newblock Tech. rep., Principia Cybernetica.
\newblock http://pcp.vub.ac.be/Papers/MediatorEvolution.pdf.

\bibitem[\protect\citeauthoryear{Heylighen}{Heylighen}{2003b}]{Heylighen2003so%
s}
{\sc Heylighen, F.} 2003b.
\newblock The science of self-organization and adaptivity.
\newblock In {\em The Encyclopedia of Life Support Systems}. EOLSS Publishers.
\newblock http://pespmc1.vub.ac.be/Papers/EOLSS-Self-Organiz.pdf.

\bibitem[\protect\citeauthoryear{Heylighen and Campbell}{Heylighen and
  Campbell}{1995}]{HeylighenCampbell1995}
{\sc Heylighen, F.} {\sc and} {\sc Campbell, D.~T.} 1995.
\newblock Selection of organization at the social level: Obstacles and
  facilitators of metasystem transitions.
\newblock {\em World Futures: the Journal of General Evolution\/}~{\em 45},
  181--212.
\newblock http://pcp.vub.ac.be/Papers/SocialMST.pdf.

\bibitem[\protect\citeauthoryear{Heylighen and Gershenson}{Heylighen and
  Gershenson}{2003}]{HeylighenGershenson2003}
{\sc Heylighen, F.} {\sc and} {\sc Gershenson, C.} 2003.
\newblock The meaning of self-organization in computing.
\newblock {\em IEEE Intelligent Systems\/}, 72--75.
\newblock http://pcp.vub.ac.be/Papers/IEEE.Self-organization.pdf.

\bibitem[\protect\citeauthoryear{Holland}{Holland}{1995}]{Holland1995}
{\sc Holland, J.~H.} 1995.
\newblock {\em Hidder Order: How Adaptation Builds Complexity}.
\newblock Helix books. Addison-Wesley.

\bibitem[\protect\citeauthoryear{Jacobson, Booch, and Rumbaugh}{Jacobson
  et~al\mbox{.}}{1999}]{JacobsonEtAl1999}
{\sc Jacobson, I.}, {\sc Booch, G.}, {\sc and} {\sc Rumbaugh, J.} 1999.
\newblock {\em The Unified Software Development Process}.
\newblock Addison-Wesley Object Technology Series. Addison-Wesley Longman
  Publishing Co., Inc., Boston, MA.

\bibitem[\protect\citeauthoryear{Jakobi}{Jakobi}{1997}]{Jakobi1997}
{\sc Jakobi, N.} 1997.
\newblock Evolutionary robotics and the radical envelope of noise hypothesis.
\newblock {\em Adaptive Behavior\/}~{\em 6,\/}~2, 325--368.

\bibitem[\protect\citeauthoryear{Jen}{Jen}{2005}]{Jen2005}
{\sc Jen, E.}, Ed. 2005.
\newblock {\em Robust Design: A Repertoire of Biological, Ecological, and
  Engineering Case Studies}.
\newblock Santa Fe Institute Studies on the Sciences of Complexity. Oxford
  University Press.
\newblock
  http://www.santafe.edu/research/publications/bookinforev/jen-info.php.

\bibitem[\protect\citeauthoryear{Jennings}{Jennings}{2000}]{Jennings2000}
{\sc Jennings, N.~R.} 2000.
\newblock On agent-based software engineering.
\newblock {\em Artificial Intelligence\/}~{\em 117,\/}~2, 277--296.
\newblock http://dx.doi.org/10.1016/S0004-3702(99)00107-1.

\bibitem[\protect\citeauthoryear{Jones, Contractor, O'Keefe, and Lu}{Jones
  et~al\mbox{.}}{1994}]{JonesEtAl1994}
{\sc Jones, P.~M.}, {\sc Contractor, N.}, {\sc O'Keefe, B.}, {\sc and} {\sc Lu,
  S. C.-Y.} 1994.
\newblock Competence models and self-organizing systems: Towards intelligent,
  evolvable, collaborative support.
\newblock In {\em Systems, Man, and Cybernetics, 1994. 'Humans, Information and
  Technology'., 1994 {IEEE} International Conference on}. Vol.~1. 367 -- 372,
  vol. 1.
\newblock http://dx.doi.org/10.1109/ICSMC.1994.399866.

\bibitem[\protect\citeauthoryear{Kaelbling, Littman, and Moore}{Kaelbling
  et~al\mbox{.}}{1996}]{KaelblingEtAl1996}
{\sc Kaelbling, L.~P.}, {\sc Littman, M.~L.}, {\sc and} {\sc Moore, A.~W.}
  1996.
\newblock Reinforcement learning: A survey.
\newblock {\em Journal of Artificial Intelligence Research\/}~{\em 4},
  237--285.
\newblock http://arxiv.org/abs/cs.AI/9605103.

\bibitem[\protect\citeauthoryear{Kauffman}{Kauffman}{1969}]{Kauffman1969}
{\sc Kauffman, S.~A.} 1969.
\newblock Metabolic stability and epigenesis in randomly constructed genetic
  nets.
\newblock {\em Journal of Theoretical Biology\/}~{\em 22}, 437--467.

\bibitem[\protect\citeauthoryear{Kauffman}{Kauffman}{1993}]{Kauffman1993}
{\sc Kauffman, S.~A.} 1993.
\newblock {\em The Origins of Order}.
\newblock Oxford University Press.

\bibitem[\protect\citeauthoryear{Kauffman}{Kauffman}{2000}]{Kauffman2000}
{\sc Kauffman, S.~A.} 2000.
\newblock {\em Investigations}.
\newblock Oxford University Press.

\bibitem[\protect\citeauthoryear{Kimura}{Kimura}{1983}]{Kimura1983}
{\sc Kimura, M.} 1983.
\newblock {\em The Neutral Theory of Molecular Evolution}.
\newblock Cambridge University Press, Cambridge.

\bibitem[\protect\citeauthoryear{Layne and Lee}{Layne and
  Lee}{2001}]{LayneLee2001}
{\sc Layne, K.} {\sc and} {\sc Lee, J.} 2001.
\newblock Developing fully functional {E-government}: A four stage model.
\newblock {\em Government Information Quarterly\/}~{\em 18}, 122--136.
\newblock http://dx.doi.org/10.1016/S0740-624X(01)00066-1.

\bibitem[\protect\citeauthoryear{Lenaerts}{Lenaerts}{2003}]{Lenaerts2003}
{\sc Lenaerts, T.} 2003.
\newblock Different levels of selection in artificial evolutionary systems:
  Analysis and simulation of selection dynamics.
\newblock Ph.D. thesis, Vrije Universiteit Brussel.
\newblock http://iridia.ulb.ac.be/\%7Etlenaert/phd/index.html.

\bibitem[\protect\citeauthoryear{Lendaris}{Lendaris}{1964}]{Lendaris1964}
{\sc Lendaris, G.~G.} 1964.
\newblock On the definition of self-organizing systems.
\newblock {\em Proceedings of the IEEE\/}.

\bibitem[\protect\citeauthoryear{Maes}{Maes}{1994}]{Maes1994}
{\sc Maes, P.} 1994.
\newblock Modeling adaptive autonomous agents.
\newblock {\em Artificial Life\/}~{\em 1,\/}~1\&2, 135 -- 162.
\newblock citeseer.ist.psu.edu/maes94modeling.html.

\bibitem[\protect\citeauthoryear{Michod}{Michod}{1997}]{Michod1997}
{\sc Michod, R.~E.} 1997.
\newblock Cooperation and conflict in the evolution of individuality. i.
  multi-level selection of the organism.
\newblock {\em American Naturalist\/}~{\em 149}, 607--645.
\newblock
  http://eebweb.arizona.edu/Michod/Downloads/Cooperation
n
0the

\bibitem[\protect\citeauthoryear{Michod}{Michod}{2003}]{Michod2003}
{\sc Michod, R.~E.} 2003.
\newblock Cooperation and conflict mediation during the origin of
  multicellularity.
\newblock In {\em Genetic and Cultural Evolution of Cooperation},
  {P.~Hammerstein}, Ed. MIT Press, Cambridge, MA, Chapter~16, 261--307.
\newblock http://eebweb.arizona.edu/Michod/Downloads/Dahlem

\bibitem[\protect\citeauthoryear{{Miller Medina}}{{Miller
  Medina}}{2005}]{MillerMedina2005}
{\sc {Miller Medina}, E.} 2005.
\newblock The state machine: Politics, ideology, and computation in {Chile},
  1964-1973.
\newblock Ph.D. thesis, MIT.
\newblock http://stuff.mit.edu/people/eden/proj.html.

\bibitem[\protect\citeauthoryear{Mitchell}{Mitchell}{1996}]{Mitchell1996}
{\sc Mitchell, M.} 1996.
\newblock {\em An Introduction to Genetic Algorithms}.
\newblock MIT Press.

\bibitem[\protect\citeauthoryear{Nicolis and Prigogine}{Nicolis and
  Prigogine}{1977}]{NicolisPrigogine1977}
{\sc Nicolis, G.} {\sc and} {\sc Prigogine, I.} 1977.
\newblock {\em Self-Organization in Non-Equilibrium Systems: From Dissipative
  Structures to Order Through Fluctuations}.
\newblock Wiley.

\bibitem[\protect\citeauthoryear{Prigogine and Herman}{Prigogine and
  Herman}{1971}]{PrigogineHerman1971}
{\sc Prigogine, I.} {\sc and} {\sc Herman, R.} 1971.
\newblock {\em Kinetic Theory of Vehicular Traffic}.
\newblock Elsevier, New York.

\bibitem[\protect\citeauthoryear{Ramamoorthy, Zhang, Fubao, and
  Ramachandran}{Ramamoorthy et~al\mbox{.}}{1993}]{RamamoorthyEtAl1993}
{\sc Ramamoorthy, P.}, {\sc Zhang, S.}, {\sc Fubao, C.}, {\sc and} {\sc
  Ramachandran, D.} 1993.
\newblock A new paradigm for the design of nonlinear dynamical systems and
  self-organizing systems.
\newblock In {\em Intelligent Control, 1993., Proceedings of the 1993 {IEEE}
  International Symposium on}. 571--576.
\newblock http://dx.doi.org/10.1109/ISIC.1993.397634.

\bibitem[\protect\citeauthoryear{Riolo, Cohen, and Axelrod}{Riolo
  et~al\mbox{.}}{2001}]{RioloEtAl2001}
{\sc Riolo, R.}, {\sc Cohen, M.~D.}, {\sc and} {\sc Axelrod, R.~M.} 2001.
\newblock Evolution of cooperation without reciprocity.
\newblock {\em Nature\/}~{\em 414}, 441--443.

\bibitem[\protect\citeauthoryear{Rojas}{Rojas}{1996}]{Rojas1996}
{\sc Rojas, R.} 1996.
\newblock {\em Neural Networks: A Systematic Introduction.}
\newblock Springer, Berlin.

\bibitem[\protect\citeauthoryear{Rosen}{Rosen}{1985}]{Rosen1985}
{\sc Rosen, R.} 1985.
\newblock {\em Anticipatory Systems: Philosophical, Mathematical and
  Methodological Foundations}.
\newblock Pergamon Press.
\newblock http://www.panmere.com/rosen/det

\bibitem[\protect\citeauthoryear{Sastry and Bodson}{Sastry and
  Bodson}{1994}]{SastryBodson1994}
{\sc Sastry, S.} {\sc and} {\sc Bodson, M.} 1989-1994.
\newblock {\em Adaptive Control: Stability, Convergence, and Robustness}.
\newblock Prentice-Hall.
\newblock http://www.ece.utah.edu/\%7Ebodson/acscr/.

\bibitem[\protect\citeauthoryear{Schreckenberg and Wolf}{Schreckenberg and
  Wolf}{1998}]{Traffic97}
{\sc Schreckenberg, M.} {\sc and} {\sc Wolf, D.~E.}, Eds. 1998.
\newblock {\em Traffic and Granular Flow '97}. Springer, Singapore.

\bibitem[\protect\citeauthoryear{Schweitzer}{Schweitzer}{2003}]{Schweitzer2003}
{\sc Schweitzer, F.} 2003.
\newblock {\em Brownian Agents and Active Particles. Collective Dynamics in the
  Natural and Social Sciences}.
\newblock Springer Series in Synergetics. Springer, Berlin.

\bibitem[\protect\citeauthoryear{Shalizi}{Shalizi}{2001}]{Shalizi2001}
{\sc Shalizi, C.~R.} 2001.
\newblock Causal architecture, complexity and self-organization in time series
  and cellular automata.
\newblock Ph.D. thesis, University of Wisconsin at Madison.
\newblock http://www.santafe.edu/projects/CompMech/papers/crs-thesis.html.

\bibitem[\protect\citeauthoryear{Simon}{Simon}{1996}]{Simon1996}
{\sc Simon, H.~A.} 1996.
\newblock {\em The Sciences of the Artificial\/}, 3rd ed.
\newblock MIT Press.

\bibitem[\protect\citeauthoryear{Sk{\aa}r and Coveney}{Sk{\aa}r and
  Coveney}{2003}]{SkarCoveney2003}
{\sc Sk{\aa}r, J.} {\sc and} {\sc Coveney, P.~V.}, Eds. 2003.
\newblock {\em Self-Organization: The Quest for the Origin and Evolution of
  Structure}. Phil. Trans. R. Soc. Lond. A 361(1807).
\newblock Proceedings of the 2002 {Nobel Symposium} on self-organization,
  http://www.pubs.royalsoc.ac.uk/phil\%5Ftrans\%5Fphys\%5Fcontent/news/selforg%
.html.

\bibitem[\protect\citeauthoryear{Steels}{Steels}{1993}]{Steels1993}
{\sc Steels, L.} 1993.
\newblock Building agents out of autonomous behavior systems.
\newblock In {\em The Artificial Life Route to Artificial Intelligence:
  Building Embodied Situated Agents}, {L.~Steels} {and} {R.~A. Brooks}, Eds.
  Lawrence Erlbaum.

\bibitem[\protect\citeauthoryear{Steels}{Steels}{1998}]{Steels1998}
{\sc Steels, L.} 1998.
\newblock Synthesising the origins of language and meaning using co-evolution,
  self-organisation and level formation.
\newblock In {\em Approaches to the Evolution of Language}, {J.~R. Hurford},
  {M.~Studdert-Kennedy}, {and} {C.~Knight}, Eds. Cambridge University Press,
  384--404.

\bibitem[\protect\citeauthoryear{{Ten Haaf}, Bikker, and Adriaanse}{{Ten Haaf}
  et~al\mbox{.}}{2002}]{tenHaafEtAl2002}
{\sc {Ten Haaf}, W.}, {\sc Bikker, H.}, {\sc and} {\sc Adriaanse, D.~J.} 2002.
\newblock {\em Fundamentals of Business Engineering and Management, A Systems
  Approach to People and Organisations}.
\newblock Delft University Press.

\bibitem[\protect\citeauthoryear{Theraulaz and Bonabeau}{Theraulaz and
  Bonabeau}{1999}]{TheraulazBonabeau1999}
{\sc Theraulaz, G.} {\sc and} {\sc Bonabeau, E.} 1999.
\newblock A brief history of stimergy.
\newblock {\em Artificial Life\/}~{\em 5,\/}~2 (Spring), 97 -- 116.
\newblock
  http://mitpress.mit.edu/catalog/item/default.asp?sid=F2804878-FE00-422F-BF58%
-2AAA736DA1C6{\&}ttype=6{\&}tid=116.

\bibitem[\protect\citeauthoryear{Turchin}{Turchin}{1977}]{Turchin1977}
{\sc Turchin, V.} 1977.
\newblock {\em The Phenomenon of Science. A Cybernetic Approach to Human
  Evolution}.
\newblock Columbia University Press, New York.
\newblock http://pespmc1.vub.ac.be/POSBOOK.html.

\bibitem[\protect\citeauthoryear{Valckenaers, {Van Brussel}, Hadeli, Bochmann,
  {Saint Germain}, and Zamfirescu}{Valckenaers
  et~al\mbox{.}}{2003}]{ValckenaersEtAl2003}
{\sc Valckenaers, P.}, {\sc {Van Brussel}, H.}, {\sc Hadeli}, {\sc Bochmann,
  O.}, {\sc {Saint Germain}, B.}, {\sc and} {\sc Zamfirescu, C.} 2003.
\newblock On the design of emergent systems: An investigation of integration
  and interoperability issues.
\newblock {\em Engineering Applications of Artificial Intelligence\/}~{\em
  16,\/}~4, 377--393.
\newblock http://dx.doi.org/10.1016/S0952-1976(03)00080-0.

\bibitem[\protect\citeauthoryear{Vaz and Varela}{Vaz and
  Varela}{1978}]{VazVarela1978}
{\sc Vaz, N.~M.} {\sc and} {\sc Varela, F.~J.} 1978.
\newblock Self and non-sense: An organism-centered approach to immunology.
\newblock {\em Medical Hypothesis\/}~{\em 4,\/}~3, 231--267.
\newblock http://dx.doi.org/10.1016/0306-9877(78)90005-1.

\bibitem[\protect\citeauthoryear{Vincent and Young}{Vincent and
  Young}{1986}]{VincentYoung1986}
{\sc Vincent, R.~A.} {\sc and} {\sc Young, C.~P.} 1986.
\newblock Self optimising traffic signal control using microprocessors - the
  {TRRL} {MOVA} strategy for isolated intersections.
\newblock {\em Traffic Engineering and Control\/}~{\em 27,\/}~7-8
  (July/August), 385--387.

\bibitem[\protect\citeauthoryear{{von Foerster}}{{von
  Foerster}}{1960}]{vonFoerster1960}
{\sc {von Foerster}, H.} 1960.
\newblock On self-organizing systems and their environments.
\newblock In {\em Self-Organizing Systems}, {M.~C. Yovitts} {and} {S.~Cameron},
  Eds. Pergamon, 31--50.

\bibitem[\protect\citeauthoryear{{von Neumann}}{{von
  Neumann}}{1956}]{vonNeumann1956}
{\sc {von Neumann}, J.} 1956.
\newblock Probabilistic logics and the synthesis of reliable organisms from
  unreliable components.
\newblock In {\em Automata Studies}, {C.~Shannon} {and} {J.~McCarthy}, Eds.
  Princeton University Press, Princeton.

\bibitem[\protect\citeauthoryear{{von Neumann}}{{von
  Neumann}}{1966}]{vonNeumann1966}
{\sc {von Neumann}, J.} 1966.
\newblock {\em The Theory of Self-Reproducing Automata}.
\newblock University of Illinois Press.
\newblock Edited by A. W. Burks.

\bibitem[\protect\citeauthoryear{Wagner}{Wagner}{2004}]{Wagner2004}
{\sc Wagner, A.} 2004.
\newblock Distributed robustness versus redundancy as causes of mutational
  robustness.
\newblock Tech. Rep. 04-06-018, Santa Fe Institute.
\newblock http://www.santafe.edu/research/publications/wpabstract/200406018.

\bibitem[\protect\citeauthoryear{Wagner}{Wagner}{2005}]{Wagner2005}
{\sc Wagner, A.} 2005.
\newblock {\em Robustness and Evolvability in Living Systems}.
\newblock Princeton University Press, Princeton, NJ.
\newblock http://www.pupress.princeton.edu/titles/8002.html.

\bibitem[\protect\citeauthoryear{Watson}{Watson}{2002}]{Watson2002}
{\sc Watson, R.~A.} 2002.
\newblock Compositional evolution: Interdisciplinary investigations in
  evolvability, modularity, and symbiosis.
\newblock Ph.D. thesis, Brandeis University.
\newblock http://demo.cs.brandeis.edu/papers/watson\%5Fthesis\%5F2002.pdf.

\bibitem[\protect\citeauthoryear{Wiener}{Wiener}{1948}]{Wiener1948}
{\sc Wiener, N.} 1948.
\newblock {\em Cybernetics; or, Control and Communication in the Animal and the
  Machine.}
\newblock Wiley and Sons, New York.

\bibitem[\protect\citeauthoryear{Wilensky}{Wilensky}{1999}]{Wilensky1999}
{\sc Wilensky, U.} 1999.
\newblock {NetLogo}.
\newblock {http://ccl.northwestern.edu/netlogo}.

\bibitem[\protect\citeauthoryear{Wilensky and Stroup}{Wilensky and
  Stroup}{2002}]{WilenskyStroup2002}
{\sc Wilensky, U.} {\sc and} {\sc Stroup, W.} 2002.
\newblock {NetLogo HubNet Gridlock} model.
\newblock {http://ccl.northwestern.edu/netlogo/models/HubNetGridlock}.

\bibitem[\protect\citeauthoryear{Wolf, Schreckenberg, and Bachem}{Wolf
  et~al\mbox{.}}{1996}]{Traffic95}
{\sc Wolf, D.~E.}, {\sc Schreckenberg, M.}, {\sc and} {\sc Bachem, A.}, Eds.
  1996.
\newblock {\em Traffic and Granular Flow '95}. World Scientific, Singapore.

\bibitem[\protect\citeauthoryear{Wooldridge}{Wooldridge}{2002}]{Wooldridge2002}
{\sc Wooldridge, M.} 2002.
\newblock {\em An Introduction to {MultiAgent} Systems}.
\newblock John Wiley and Sons, Chichester, England.
\newblock http://www.csc.liv.ac.uk/\%7Emjw/pubs/imas/.

\bibitem[\protect\citeauthoryear{Wooldridge and Jennings}{Wooldridge and
  Jennings}{1995}]{WooldridgeJennings1995}
{\sc Wooldridge, M.} {\sc and} {\sc Jennings, N. R.~.} 1995.
\newblock Intelligent agents: Theory and practice.
\newblock {\em The Knowledge Engineering Review\/}~{\em 10,\/}~2, 115--152.
\newblock http://www.ecs.soton.ac.uk/\%7Enrj/download-files/KE-REVIEW-95.ps.

\bibitem[\protect\citeauthoryear{Wooldridge, Jennings, and Kinny}{Wooldridge
  et~al\mbox{.}}{2000}]{WooldridgeEtAl2000}
{\sc Wooldridge, M.}, {\sc Jennings, N.~R.}, {\sc and} {\sc Kinny, D.} 2000.
\newblock The {Gaia} methodology for agent-oriented analysis and design.
\newblock {\em Journal of Autonomous Agents and Multi-Agent Systems\/}~{\em
  3,\/}~3, 285--312.
\newblock http://www.ecs.soton.ac.uk/\%7Enrj/download-files/jaamas2000.pdf.

\bibitem[\protect\citeauthoryear{Zambonelli, Jennings, and
  Wooldridge}{Zambonelli et~al\mbox{.}}{2003}]{ZambonelliEtAl2003}
{\sc Zambonelli, F.}, {\sc Jennings, N.~R.}, {\sc and} {\sc Wooldridge, M.}
  2003.
\newblock Developing multiagent systems: The {Gaia} methodology.
\newblock {\em ACM Trans on Software Engineering and Methodology\/}~{\em
  12,\/}~3, 317--370.
\newblock http://www.ecs.soton.ac.uk/\%7Enrj/download-files/tosem03.pdf.

\bibitem[\protect\citeauthoryear{Zambonelli and Rana}{Zambonelli and
  Rana}{2005}]{ZambonelliRana2005}
{\sc Zambonelli, F.} {\sc and} {\sc Rana, O.~F.} 2005.
\newblock Self-organization in distributed systems engineering: Introduction to
  the special issue.
\newblock {\em Systems, Man and Cybernetics, Part A, IEEE Transactions
  on\/}~{\em 35,\/}~3 (May), 313 -- 315.
\newblock http://dx.doi.org/10.1109/TSMCA.2006.846372.

\end{thebibliography}
}

\end{document}